\documentclass[10.5pt]{article}

\usepackage{color,graphicx}
\usepackage{young}
\usepackage[vcentermath]{youngtab}
\usepackage{amsmath,amssymb,graphicx}
\usepackage{hyperref}
\definecolor{darkred}{rgb}{0.65,0.15,0}
\hypersetup{pdfborder={0 0 0},colorlinks=true,urlcolor=darkred,citecolor=blue,linkcolor=darkred,linktocpage=true}

\usepackage{cite}
\usepackage{amsmath}
\usepackage{amsfonts}
\usepackage{amssymb}
\usepackage{graphicx}%
\usepackage{amsthm}
\usepackage{mathrsfs}
\usepackage[T1]{fontenc}
\usepackage{enumerate}
\usepackage{color}

\theoremstyle{definition}

\def\4diml{four-dimensional}

\def\-1{^{-1}}

\newcommand{\A}{\mathscr{A}}

\newcommand{\C}{\mathscr{C}}
\newcommand{\M}{\mathscr{M}}
\newcommand{\D}{\mathscr{D}}
\newcommand{\G}{\mathscr{G}}

\newcommand{\tG}{\widetilde{\mathscr{G}}}

\makeatletter

\@addtoreset{equation}{section}
\makeatother

\begin{document}

\thispagestyle{empty}

\vspace{5mm}

\begin{center}
{\LARGE \bf  Classification of Conformal Supersymmetric \\[1mm]  Deformations of Schwarzschild Spacetime
via Super \\[2mm]  Poisson-Lie T-duality/Plurality}

\vspace{15mm}

\normalsize
{\large  Ali Eghbali\footnote{Corresponding author: eghbali978@gmail.com}, Meysam Hosseinpour-Sadid\footnote{meysam.hs.az@gmail.com},
Adel Rezaei-Aghdam\footnote{rezaei-a@azaruniv.ac.ir}}\\

\vspace{2mm}
{\small \em Department of Physics, Faculty of Basic Sciences,\\
Azarbaijan Shahid Madani University, 53714-161, Tabriz, Iran}\\
\vspace{4mm}

\vspace{7mm}

\vspace{6mm}

\begin{tabular}{p{12cm}}
{\small
This paper investigates conformal supersymmetric deformations of Schwarzschild spacetime through the framework of super Poisson-Lie T-duality/plurality.
We first study super non-Abelian T-duality in $\sigma$-models featuring a Schwarzschild metric coupled to two fermionic fields.
This is realized by describing the original four-dimensional geometry and its dual counterpart via semi-Abelian Drinfeld superdoubles generated by the Lie superalgebras $({\C}^3 +{\A})$, ${\C}^3 \oplus {\A}_{1,1}$ and $(2{ \A}_{1,1}+2{ \A})^0$, supplemented by two spectator fields.
We enforce the vanishing of one-loop beta-function equations for both original and dual models to guarantee UV finiteness at
quantum level;
this procedure necessitates the inclusion of the dilaton field alongside the metric and the
Kalb-Ramond field ($B$-field) already present in the classical action.
We further study the curvature invariants, singularity structure, and superisometries, and compare the
resulting supergeometries, with particular emphasis on their fermionic parts and $B$-fields.
Starting from the decompositions of semi-Abelian Drinfeld superdoubles associated with the $({\C}^3 +{\A})$ and ${\C}^3 \oplus {\A}_{1,1}$ Lie superalgebras,
we derive the conformal duality/plurality chains for four-dimensional string backgrounds, characterized by a Schwarzschild metric coupled to two fermions.
Our results span an interesting spectrum of super Poisson-Lie T-dual $\sigma$-models described by Lie superalgebras
with two bosonic and two fermionic generators. These are all non-trivial and interesting examples of super Poisson-Lie T-dual models which
help in the intent of providing a general classification of four-dimensional geometries describing
supergravity backgrounds.
}
\end{tabular}
\vspace{-1mm}
\end{center}

\begin{center}

\end{center}
\setcounter{page}{1}
\newpage
\tableofcontents

 \vspace{5mm}

\section{Introduction}
\label{Sec.I}

The existence of target space duality (T-duality) is one of the most interesting features discovered in non-trivial string backgrounds.
T-duality is a general property of string vacua that possess at least one isometry,
and connects seemingly different backgrounds in which the strings can propagate \cite{{Buscher1},{Buscher2}}.
Buscher showed that \cite{{Buscher1},{Buscher2}} Abelian T-duality is possible when the group of isometries of the target manifold is
Abelian. He realized that \cite{{Buscher1},{Buscher2}} duality symmetries are a property of all string vacua with Abelian isometries.
It goes without saying that before this, duality symmetries were originally discovered for toroidal compactifications of closed string theories \cite{{kikkawa},{Sakai}}.
Subsequently, the invariance of the partition function under a
duality transformation was shown for the corresponding $\sigma$-model.
Then,  de la Ossa and Quevedo \cite{de.laOssa} generalized Buscher's duality to backgrounds with non-Abelian isometries.
They understood that this new duality transformation maps spaces with non-Abelian isometries to spaces that may
have no isometries at all. Accordingly, it can be suggested that the duality symmetries in string theories need
to be understood in a more general context without regard to the existence of continuous
isometries on the target space. Various aspects of the non-Abelian T-duality along with some examples were discussed in \cite{{Rabinovici},{Giveon},{Alvarez1},{Alvarez2}}.
One of the first physically interesting examples given for the non-Abelian T-duality formulation was the Schwarzschild metric.
The Schwarzschild metric as the most general spherically symmetric vacuum solution of the Einstein field equations is given by \cite{Schwarzschild}
\begin{eqnarray}
ds^2=(1-\frac{2M}{r})^{-1} ~ dr^2   -(1-\frac{2M}{r}) ~ {d t^2} + r^{2} \big({d \theta}^2 + \sin^2 \theta ~ {d \varphi^2 }\big).\label{1.1}
\end{eqnarray}
Indeed, this geometry satisfies the standard supergravity equations with a constant dilaton and zero $B$-field.
In \cite{de.laOssa}, the non-Abelian dual background of the Schwarzschild metric
whose isometry group is $SO(3)\times \{$time translations$\}$, was obtained and explicitly shown to be new solution
of the standard supergravity equations.
The new background possessed no continuous isometries, with the exception of time translations, and exhibited a naked singularity (see, also, \cite{hewson}).
Note that in the duality scheme invited by de la Ossa and Quevedo, the dual model might not have isometries anymore, so it was not possible to
perform the duality transformation again to obtain the original model.
A major step to clarify this problem was made by Klimcik and Severa \cite{{Klim1},{Klim2}}.
They replaced the symmetry condition with a weaker one.
It deals with $\sigma$-models based on two Lie groups and the corresponding duality transformations are known as Poisson-Lie T-duality.
In this kind of the duality, the model is not required to be symmetric, but there is still an action of a group $G$ on the
manifold and the Noetherian currents associated to this action are required
to be integrable.
Poisson-Lie T-duality proposed by Klimcik and Severa was extended to a supermanifold case, especially a Lie supergroup one,
and was then called super Poisson-Lie T-duality \cite{ER2} (see, also, \cite{ER5}).
In recent years, we have witnessed further interest in super Poisson-Lie T-duality, driven by $\sigma$-models based on Lie supergroups
\cite{jan.Vysoky,ER7,ER8,Eghbali,Bielli1,Bielli2,falk.Hassler,ER15}.
In \cite{falk.Hassler},  to describe generalized dualities, it is discussed how to
employ superspace double field theory, involving a generalized supervielbein,
an element of $OSp(D, D|2s)$.
A small but important difference in the scheme given in \cite{falk.Hassler} compared to ours
is that they do not require an invertible supermetric, which is important for applications of the Green-Schwarz superstring.
In a recent study, we investigated super (non-)Abelian T-duality and
super T-plurality\footnote{Poisson-Lie T-plurality introduced by von Unge \cite{vonUnge} to include the
case of Drinfeld doubles which can be decomposed into Lie bialgebras in more than one way.
It was then generalized to a Lie supergroup case to present the conformal duality chains of
$2+1$-dimensional cosmological string backgrounds coupling with two fermionic fields \cite{Eghbali3}.
An interesting study has also shown Poisson-Lie T-plurality on Wess-Zumino-Witten backgrounds\cite{sakatani1} (see, also, \cite{I.Petr}).}
 for the BTZ metric coupled to two fermionic fields \cite{Meysam}.
By constructing various examples of mutually T-dual $\sigma$-models via semi-Abelian superdoubles, we demonstrated that distinct Drinfeld superdoubles yield different couplings between the BTZ coordinates and the fermionic sectors.
Furthermore, for a specific Drinfeld superdouble, our analysis was extended to the construction of a chain of models dual to the BTZ vacuum plus two fermions.
This chain is associated with isomorphic Manin supertriples, arising from the multiple decompositions available for the given superdouble.
The main purpose of the present paper is twofold.

\begin{itemize}

\item  Firstly, by using a certain parametrization of the
$(2|2)$-dimensional Lie supergroups $({C}^3 +{A})$, ${C}^3 \otimes {A}_{1,1}$
and $(2{A}_{1,1}+2{A})^0$ \cite{B,ER6} and by a suitable choice of spectator-dependent matrices
we construct the original $\sigma$-models including the Schwarzschild metric plus two fermions
and a non-trivial $B$-field.
By solving the vanishing beta-function equations at the one-loop level, several remarkable properties of the original models emerge.
Then, by applying super non-Abelian T-duality (here as super Poisson-Lie T-duality on a semi-Abelian superdouble)
we construct three non-Abelian dual pairs for the Schwarzschild metric coupled to two fermionic fields.
The dual backgrounds are supported by some non-trivial $H$-fluxes and the corresponding metrics
contain the true singularities in both radial and angular regions.
We thus find some solutions of the string background equations, which are not black hole, but have naked singularities
and are dual to the Schwarzschild metric coupled to two fermionic fields.

\item  Secondly, starting from the decomposition of semi-Abelian Drinfeld superdoubles generated by the $({\C}^3 +{\A})$ and $
{\C}^3 \oplus {\A}_{1,1}$ Lie superbialgebras
we study the super Poisson-Lie T-plurality of the Schwarzschild spacetime coupled to two fermionic fields, in such a way that
the conformal duality/plurality chains of four-dimensional string backgrounds are obtained.
However, our findings as a spectrum of super Poisson-Lie T-dual $\sigma$-models described by the $(2|2)$-dimensional Lie superalgebras
are interesting in themselves, but at a constructive level, can prompt many new insights into (generalized) supergravity and manifestly have interesting mathematical relationships with double field theory.

\end{itemize}

The organization of this paper is as follows:
In Sec. \ref{Sec.II} we recall the properties of $\mathbb{Z}_2$-graded vector
space and outline the basic definitions related to Lie superalgebras. We also review aspects of super Poisson-Lie
T-duality \cite{ER2, ER5} that are directly relevant to our context.
In Sec. \ref{Sec.III}, we construct supersymmetric deformations of the Schwarzschild spacetime coupled to two fermionic fields
and obtain their super non-Abelian T-dual backgrounds. The backgrounds are constructed on three semi-Abelian Drinfeld superdoubles, namely
$({\C}^3 \oplus {\A}_{1,1} , {\cal I}_{_{(2|2)}})$, $(({\C}^3 + \A) , {\cal I}_{_{(2|2)}})$,
and $((2{\A}_{1,1}+2{\A})^0 , {\cal I}_{_{(2|2)}})$.
The resulting backgrounds are analyzed in terms of their conformal properties, curvature invariants, singularity structure, and superisometries.
We also compare the three original and dual supergeometries, with particular attention to the differences in their fermionic parts and $B$-fields.
In Sec. \ref{Sec.IV}, we extend the analysis to super Poisson-Lie T-plurality by considering alternative Manin supertriple decompositions of the corresponding Drinfeld superdoubles.
Subsection \ref{IV.1} briefly reviews the formulation of super Poisson-Lie T-plurality,
while subsections \ref{IV.2} and \ref{IV.3} construct further conformal backgrounds associated with the $C^3 \otimes A_{1,1}$ and $(C^3+A)$ Lie supergroups, respectively.
The one-loop conformal invariance of these backgrounds is then verified.
Finally, Sec. \ref{Sec.V} presents the concluding remarks and discusses possible directions for future work.

\section{Review of super Poisson-Lie T-duality in the presence of spectator fields}
\label{Sec.II}
In this section, we shall give a concise review of the super Poisson-Lie T-duality on supermanifolds \cite{ER2,ER5}.
Before proceeding, let us recall the properties of $\mathbb{Z}_2$-graded vector
space and also some definitions related to Lie superalgebras \cite{N.A}.

\subsubsection*{{\it \underline{Notations and basic definitions}}}

A supervector space $V$ is a ${\mathbb{Z}}_{2}$-graded vector space, a vector space as ${V}= { V}_{0} \oplus {V}_{1}$ over an arbitrary field $\mathbb{F}$.
Elements of the $0$-graded part ${V}_{0}$
are called even (bosonic) and elements of ${V}_{1}$ are called odd (fermionic). If $x \in V$ is homogeneous, we will denote
by $|x| \in  \{0, 1\}$ the graded degree of $x \in V$, namely,
$|x|=0$ for any $x \in {V}_{_0}$, while $|x|=1$ for any $x \in {V}_{_1}$.
A superalgebra (associative or not) is a ${\mathbb{Z}}_{2}$-graded algebra. The grading thus satisfies
$|xy| = |x| + |y|$ in ${\mathbb{Z}}_{2}$. Thus it is a supervector space with a compatible multiplicative.

A Lie superalgebra is a supervector space ${\G}$ with a graded non-associative algebra, thus admitting the decomposition ${\G} ={\G}_{0} \oplus {\G}_{1}$.
There is defined on ${\G}$ a {\it multiplication}, denoted by $[. , .]: {\G} \otimes {\G} \rightarrow {\G}$ satisfying the requirements of super anti-symmetry
and super Jacobi identity, respectively\footnote{From now on, we adopt the notation introduced by Dewitt \cite{D}. According to \cite{D}, grading indices are identified by the exponent of $(-1)$;
in fact, one may use $(-1)^x$ instead of $(-1)^{|x|}$, where $(-1)^x$ is $1$ or $-1$ for even or odd elements, respectively.} \cite{N.A, ER1},
\begin{align}\label{2.1}
&[x , y] =-(-1)^{|x||y|}~ [y , x],\nonumber\\
&[x , [y , z]] + (-1)^{|x|(|y|+|z|)}~ [y , [z , x]] + (-1)^{|z|(|x|+|y|)}~ [z , [x , y]]=0,
\end{align}
for homogeneous $x, y, z$.
One says that the Lie superalgebra ${\G}$ is of the superdimension $(m | n)$ iff $\dim{\G}_{0} = m$
and $\dim{\G}_{1} = n$. It may be finite or infinite. Let $\G$ be a finite dimensional Lie superalgebra, and consider its dual $\G^\ast$.
By definition, an element $x^\ast \in \G^\ast$ is a linear functional on $\G$, i.e., $x^\ast(y) = \big<x^\ast , y\big>$ for all $y \in \G$.

To evade the problems with definitions of supergroups we shall define the Drinfeld superdouble only on the algebraic level.
The Drinfeld superdouble $D$ is defined as a Lie supergroup whose Lie superalgebra, ${\D}$, equipped
by a supersymmetric ad-invariant non-degenerate bilinear form
$\big<. , .\big>$ can be decomposed into a pair of maximally isotropic sub-superalgebras ${\G}$ and ${\tilde {\G}}$,
and ${\D}$ as a vector superspace is ${\D} ={\G}\oplus {\tilde {\G}}$.
This ordered triple of Lie superalgebras $({\D}, {\G}, {\tilde {\G}})$ is called {\it Manin supertriple}.
Also, the pair of $({\G} , {\tilde {\G}})$ is called {\it Lie superbialgebra} \cite{N.A,J.z,ER1}.
The dimension of sub-superalgebras have to be equal. We furthermore consider $G$ and $\tilde G$ as a pair of maximally isotropic sub-supergroups
corresponding to the $\G$ and $\tilde \G$.
The maximally isotropic nature of the sub-superalgebras requires that bases
$\{T_{_a}\} \in \G$  and $\{{\tilde T}^a\} \in {\tilde \G}$, $a = 1, ..., \dim G$ satisfy the following conditions
\begin{eqnarray}\label{2.2}
\big<{T_{_a}} , {T_{_b}}\big> =0,~~~\big<{\tilde T}^{^a}  , {\tilde T}^{^b}\big> =0,~~~~
{{\delta}^{^b}}_{a} = \big<{\tilde T}^{^b} , {T_{_a}}\big> = (-1)^{{ab}}  \big<{T_{_a}} , {\tilde T}^{^b}\big>.
\end{eqnarray}
Note that ${\tilde {\G}}$ is considered to be a dual ${{\G}^\ast}$ to $\G$. To define the Lie super cobracket in $\G$ one may use
the Lie superbracket in ${\tilde {\G}}$.
The Lie superbracket on $\tilde \G$ defines a Lie superbracket on $\G^{\ast}$.
Furthermore, to see that it defines a Lie superbialgebra structure on $\G$,
one may apply the super Jacobi identity on $\D$ and the invariance of the scalar product.
Thus, there is a one-to-one correspondence between Lie superbialgebra $({\bf \G},{\bf
\G}^\ast)$ and Manin supertriple $(\D , \G , \tilde{\G})$ with $\tilde{\G} \cong  {\G}^\ast$ \cite{Kosmann}.
Due to the ad-invariance of the bilinear form $\big<. , .\big>$, the algebraic structure of ${\D}$ is determined by
its maximally isotropic sub-superalgebras.
In the bases $\{T_{_a}\}$  and $\{{\tilde T}^a\}$, the Lie superbrackets are given by \cite{ER1}
\begin{eqnarray}\label{2.3}
[{T_{_a}} , {T_{_b}}] &=& {f^c}_{_{ab}} ~{T_{_c}},~~~~~~~
[{\tilde T}^{^a} , {\tilde T}^{^b}] = {{\tilde f}^{ab}}_{\;   \; c} ~{\tilde T}^{^c},\nonumber\\
{[{T_{_a}} , {\tilde T}^{^b}]} &=& (-1)^{^b} {{{\tilde f}^{bc}}}_{\; \;a}~ {T_{_c}} + (-1)^{^a} {f^b}_{_{ca}} ~{\tilde T}^{^c}.
\end{eqnarray}
The Lie superalgebra structure defined by relation \eqref{2.3} is called the \emph{Drinfeld superdouble} ${\D}$.
The super Jacobi identity of Lie superalgebra ${\D}$ imposes the following relation over the structure
constants of Lie superalgebras  ${\G}$ and ${\tilde {\G}}$ \cite{ER1,ER2}
\begin{eqnarray}\label{2.4}
{f^d}_{bc}{\tilde{f}^{ae}}_{\; \; \; \; d}=
{f^a}_{dc}{\tilde{f}^{de}}_{\; \; \; \; \; b} +
{f^e}_{bd}{\tilde{f}^{ad}}_{\; \; \; \; \; c}+ (-1)^{be}
{f^a}_{bd}{\tilde{f}^{de}}_{\; \; \; \; \; c}+ (-1)^{ac}
{f^e}_{dc}{\tilde{f}^{ad}}_{\; \; \; \; \; b}.
\end{eqnarray}

\subsubsection*{{\it \underline{Super Poisson-Lie symmetry}}}

In what follows we shall consider a non-linear $\sigma$-model on a two-dimensional curved surface ${\Sigma}$ in the
$(d_{_B}|d_{_F})$-dimensional supermanifold $\M$ with
the supersymmetric metric $G_{_{MN}}$ and super anti-symmetric Kalb-Ramond field $B_{_{MN}}$. The action of model in lightcone coordinates
$\sigma^{\pm} = (\tau \pm \sigma)/2$ is given by
\begin{eqnarray}
S= \frac{1}{2}\int_{_{\Sigma}}\!d\sigma^+  d\sigma^- ~ (-1)^{^M} \partial_{_+} X^{^M}
\big({G}_{_{MN}}(X)+{B}_{_{MN }}(X)\big) \partial_{_-} X^{^N},\label{2.5}
\end{eqnarray}
where $\partial_{_\pm}$ are the derivatives with respect to the lightcone variables, and $X^{^M}$
are the supercoordinates on $\M$, which include the bosonic coordinates $x^i$, $i = 0, ..., d_{_B}$-1 and
the fermionic ones $\Theta^\alpha$, $\alpha = 1, ..., d_{_F}$ and thus the labels $M$ and $N$ run over $(i, \alpha)$.
It  seems  to  be  of  interest to define the line element $ds^2$ and $B$-field corresponding to the above
action in the coordinate basis. They are, respectively, read
\begin{eqnarray}
ds^2 =(-1)^{^{MN}} ~G_{_{MN}} dX^{^M}~dX^{^N},~~~~
B = \frac{1}{2} (-1)^{^{MN}}~B_{_{MN}} ~ dX^{^M} \wedge dX^{^N}.\label{2.6}
\end{eqnarray}

Let us suppose now that there is a Lie supergroup $G$ acting freely on  $\M$ from right.
If the Noether's current one-forms corresponding to the right action of $G$ on
$\M$ are not closed and instead satisfy the super Maurer-Cartan equation \cite{D} on the extremal surfaces,
one then says that the $\sigma$-model \eqref{2.5} has the {\it super Poisson-Lie symmetry} with respect to the dual
Lie supergroup ${\tilde G}$ (with the same superdimension $G$).
It is the condition of dualizability of $\sigma$-model on the level of the Lagrangian, which is
given by the formula \cite{ER2}
\begin{eqnarray}\label{2.7}
{\cal L}_{_{V_{_a}}}{\cal E}_{_{MN}}\;=\;(-1)^{^{a+Ma+Rc+P}}~{{\tilde f}^{bc}}_{~~a}~ {\cal E}_{_{MR}}~ {V_{_c}}^{^{~R}}~{V_{_b}}^{^{~P}}~
{\cal E}_{_{PN}},
\end{eqnarray}
where the tensor field ${\cal E}_{_{MN}}$ can be understood as a sum of the $G_{_{MN}}$ and $B_{_{MN}}$. Moreover,
${\cal L}_{_{V_{_a}}}$ stands for the Lie derivative corresponding to the left-invariant supervector
fields ${V_{_a}}= {V_{_a}}^{^M} {{{\overrightarrow \partial}/{\partial{X^{^M}}}
}}$ (defined with left derivative) satisfying $[{V_{_a}} , {V_{_b}}]= {f^c}_{ab} ~{V_{_c}}$, and
${{\tilde f}^{bc}}_{~~a}$ are the structure constants of the dual Lie superalgebra $\tilde \G$ of $\tilde G$.
Note that the integrability condition on the Lie derivative,
$[{\cal L}_{_{V_{_a}}} , {\cal L}_{_{V_{_b}}}]= {\cal L}_{_{[V_{_a} , V_{_b}]}}$,
then implies the mixed super Jacobi identities \eqref{2.4} showing that this construction leads naturally to
the Drinfeld superdouble.

\subsubsection*{{\it \underline{Super Poisson-Lie T-dual $\sigma$-models with the spectator fields}}}

In the rest of this section we will briefly review, by following the conventions of \cite{ER5}
(see, also, \cite{{Klim1},{Klim2},{Sfetsos1},{Sfetsos2}}),
some facts about super Poisson-Lie T-duality in the presence of the spectator fields, which are necessary for our constructions.

Let us now consider a target supermanifold $\M \approx O \times G$ with the $d$ field variables $X^{^{M}} = (x^{\mu} , y^i)$,
where $x^\mu$'s, $\mu = 1, \cdots, \dim G$ are  chosen to parametrize an element $g$ of a Lie supergroup $G$ (with Lie superalgebra $\G$),
whereas the rest are the so-called {\it spectators} and will be denoted by $y^i$, $i = 1, \cdots,  d-\dim G$.
These are the coordinates labeling the orbit $O$ of $G$ in ${\M}$. Note that the Lie supergroup $G$ acts
freely from right on $\M $.  We also introduce the basis $\{T_{_a}\},$~$a= 1, \cdots$, dim $G$ of Lie superalgebra $\G$ of $G$,
with  ${{R_{_\pm}}}^{\hspace{-1.5mm} a} =(\partial_{_{\pm}} g  ~g^{-1} )^a=\partial_{_\pm} x^{{\mu}}
\; {{_{_\mu}} R}^{^a}$ denoting the corresponding components of the right-invariant super one-forms.
Similarly, we introduce the dual target supermanifold ${\tilde \M} \approx O \times \tilde G$ with the field variables
${\tilde X}^{^{M}} =({\tilde x}^{\mu} , y^i)$, where
${\tilde x}^{\mu}$'s, $\mu = 1, \cdots, \dim {G}$ parameterize a different supergroup ${\tilde G}$, whose dimension is, however,
equal to that of $G$, and the rest of the variables are the same $y^i$'s introduced in ${\M}$.
Accordingly, we introduce a different set of bases $\{{\tilde T}^a \}$ of the Lie superalgebra ${\tilde \G}$ with
$a = 1, \cdots, \dim {\tilde G}$, and the corresponding right-invariant super one-forms defined as
 ${{\tilde  R_{_\pm a}}}=(\partial_{\pm}{\tilde g} ~{\tilde g}^{-1})_a
= (-1)^\mu \partial_{\pm} {\tilde x}^{\mu} {\tilde R}_{\mu a}$.
The two-dimensional ${\sigma}$-model actions related by super Poisson-Lie T-duality, in the presence of spectator fields,
are given by\footnote{Note that here we do not require any isometry associated with the Lie supergroups $G$ and $\tilde G$. The $\sigma$-models \eqref{2.8} and \eqref{2.9} will be dual to each other in the sense of super Poisson-Lie
T-duality if the associated Lie superalgebras $\G$ and ${\tilde \G}$
form a Drinfeld superdouble which can be decomposed.}
\begin{eqnarray}
S &=& \frac{1}{2} \int_{_{\Sigma}} d\sigma^{+}  d\sigma^{-} \Big[(-1)^a  R_{+}^a~{{E}_{_{ab}}}(g, y^i)~
 R_{-}^b + (-1)^a  R_{+}^a~ \phi^{{(1)}}_{a j}(g, y^i) \partial_{-} y^{j}\nonumber\\
&&~~~~~~~~~~~~~~~~~~~~~~~~~~  + (-1)^i \partial_{+} y^{i} \phi^{{(2)}}_{i b}(g, y^i)  R_{-}^b + (-1)^i~ \partial_{+} y^{i} \phi_{_{ij}}(g, y^i) \partial_{-} y^{j} \Big],\label{2.8}\\
\tilde S &=& \frac{1}{2} \int_{_{\Sigma}} d\sigma^{+}  d\sigma^{-}\Big[(-1)^b {\tilde R}_{+_{a}} {{{\tilde E}}^{{ab}}}({\tilde g}, {y}^{i})~
{\tilde R}_{-_{b}}+{\tilde R}_{+_{a}} {\tilde \phi}^{\hspace{0mm}{(1)^{ a}}}_{~~~~j}({\tilde g}, {y}^{i}) ~\partial_{-} y^{j}\nonumber\\
&&~~~~~~~~~~~~~~~~~~~~~~~~ +(-1)^{i+b} \partial_{+} y^{i} {\tilde \phi}^{\hspace{0mm}{(2)^{ b}}}_{i}({\tilde g}, {y}^{i})  ~{\tilde R}_{-_{b}}
+ (-1)^i \partial_{+} y^{i} {\tilde \phi}_{_{ij}}({\tilde g}, {y}^{i}) \partial_{-} y^{j}\Big].\label{2.9}
\end{eqnarray}
In the first action, the couplings ${{E}_{_{ab}}}, \phi^{{(1)}}_{a j}, \phi^{{(2)}}_{i b} $ and $\phi_{_{ij}}$ may depend on all variables $x^\mu$ and $y^i$. They are restricted to be of the form \cite{ER5, Sfetsos1, Sfetsos2}
\begin{eqnarray}\label{2.10}
{E_{_{ab}}} &=& {\big(E^{{-1}}_{0}+ \Pi\big)_{{ab}}}^{\hspace{-3mm}-1},~~~~~~~~~~~~~~~
\phi^{{(1)}}_{a i} = (-1)^c~{E_{_{ab}}}~({E^{{-1}}_{0}})^{bc}~{F_{~ci}}^{^{\hspace{-2mm}(1)}},\nonumber\\
\phi^{{(2)}}_{i b}&=& (-1)^c~{F_{~ia}}^{^{\hspace{-2.5mm}(2)}}~ ({E^{{-1}}_{0}})^{ac}~{E_{_{cb}}},~~
\phi_{_{ij}} = F_{_{ij}} - (-1)^{e+c}~{F_{~ia}}^{^{\hspace{-2.5mm}(2)}}~{\Pi}^{ac} {E_{_{cd}}} ~({E^{{-1}}_{0}})^{de}~{F_{~ej}}^{^{\hspace{-2mm}(1)}},~~~
\end{eqnarray}
where the new couplings $E_{0}, F^{^{(1)}}, F^{^{(2)}}$ and $F$ may be at most functions of the
variables $y^{i}$ only. In the above equation, $\Pi(g)$ defined as $\Pi^{^{ab}}(g) =(-1)^c  ~b^{^{ac}}(g)~ (a^{-1})_{_{c}}^{^{~b}}(g)$ is called
the super Poisson structure, where the matrices $a_{_{a}}^{{~b}}(g)$ and $b^{^{ab}}(g)$ are constructed using\footnote{``st'' denotes supertransposition \cite{D}.}
\begin{eqnarray}\label{2.11}
g^{-1} T_{{_a}}~ g &=& (-1)^b ~a_{_{a}}^{{~b}}(g) ~ T_{{_b}},\nonumber\\
g^{-1} {\tilde T}^{{^a}} g &=&(-1)^c~ b^{^{ac}}(g)~ T_{{_c}}+{(a^{-st})^{a}}_{{b}}(g)~{\tilde T}^{{^b}}.
\end{eqnarray}
It only remains to relate the dual action couplings to those of the original one.
Their relationship is given as follows:
\begin{eqnarray}
{{\tilde E}^{^{ab}}} &=& {\big(E_{0}+ {\tilde \Pi}\big)^{\hspace{-1mm}-1}}^{ab},~~~~~~~~~~
{\tilde \phi}^{\hspace{0mm}{(1)^{ a}}}_{~~~j} = (-1)^b~ {{\tilde E}^{^{ab}}}~{F_{~bj}}^{^{\hspace{-2.5mm}(1)}},\nonumber\\
{\tilde \phi}^{\hspace{0mm}{(2)^{ b}}}_{i} &=& - {F_{~ia}}^{^{\hspace{-2.75mm}(2)}} ~{{\tilde E}^{^{ab}}},~~~~~~~~~~~~~~~
{\tilde \phi}_{_{ij}} = F_{_{ij}}- (-1)^{b}~ {F_{~ia}}^{^{\hspace{-2.75mm}(2)}} ~{{\tilde E}^{^{ab}}}~{F_{~bj}}^{^{\hspace{-2mm}(1)}},\label{2.12}
\end{eqnarray}
where ${\tilde \Pi} (\tilde g)$ is defined similar to ${\Pi}(g)$ by just replacing the untilded symbols by tilded ones.
Note that the super non-Abelian duality is recovered if $\tilde G$ is Abelian or in other words,  $\tilde f^{ab}_{~~~c}=0$,
in which case we would be dealing with a semi-Abelian superdouble.
Then, it follows from \eqref{2.11} that $ b(g)=0 $, then, $\Pi(g)=0 $; consequently, for action \eqref{2.8} we will have $E=E_{0}, \phi^{{(1)}}=F^{^{(1)}}, \phi^{{(2)}}=F^{^{(2)}}$ and $\phi=F$.
In this way, the dual right-invariant super one-forms are written as ${\tilde R}_{\pm_{a}}=\partial_{\pm} {{\tilde x}_a}$,
where the ${{\tilde x}_a}$'s parametrize the supergroup element $\tilde g \in \tilde G$.
Furthermore, using the dual version of \eqref{2.11} we find that ${{\tilde a}^a}_{_{~}b} ({\tilde g})={{\delta}^a}_{_b}$,
then, ${\tilde \Pi}_{_{ab}} ({\tilde g})={\tilde b}_{_{ab}}(\tilde g)$.

\subsubsection*{{\it \underline{Conformal invariance of the $\sigma$-model at the one-loop level}}}

In the $\sigma$-model context, the ultraviolate finiteness of the quantum version of the model is guaranteed
by the conformal invariance of the model.
To achieve this invariance at the one-loop level, Fradkin and Tseytlin \cite{FRADKIN} suggested that one should add to the
Lagrangian of action \eqref{2.5} the renormalizable, but not Weyl invariant, term containing the so-called dilaton field.
The dilaton field $\Phi$ can be understood as an additional function on $\M$
that defines the quantum non-linear $\sigma$-model and couples to scalar curvature of
the worldsheet. The conformal invariance of the
model is guaranteed by vanishing of the so-called beta-function \cite{Callan}. At the one-loop level
the equations for the vanishing of the beta-function on a supermanifold $\M$ are given by \cite{ER5}
\begin{eqnarray}
&&{\cal R}_{_{MN}}+\frac{1}{4}H_{_{MQP}} {H^{^{PQ}}}_{_{N}}+2{\overrightarrow{\nabla}_{_M}}
{\overrightarrow{\nabla}_{_N}} \Phi ~=~0,\label{2.13}\\
&&(-1)^{^{P}} {\overrightarrow{\nabla}}^{P}\big(e^{^{-2 \Phi}} H_{_{PMN}}\big)
~=~0,\label{2.14}\\
&&4 \Lambda-{\cal R}-\frac{1}{12} H_{_{MNP}}H^{^{PNM}} +4{\overrightarrow{\nabla}_{_M}} {\Phi} \overrightarrow{\nabla}^{^M} {\Phi}
 - 4 {\overrightarrow{\nabla}_{_M}} \overrightarrow{\nabla}^{^M} {\Phi}=0,\label{2.15}
\end{eqnarray}
where the covariant derivatives ${\overrightarrow{\nabla}_{_M}}$, Ricci tensor ${\cal R}_{_{MN}}$
and scalar curvature ${\cal R}$ are
calculated from the metric $~G_{_{MN}}$ that is also used for lowering and raising indices.
$H_{_{MNP}}$ is the field strength corresponding to the $B$-field which is defined by\footnote{Here we have used the left partial differentiation to define $H=dB$. The relation between the left partial differentiation
and right one is detailed in \cite{D}.}
\begin{eqnarray}\label{2.16}
H_{_{MNP}} =(-1)^{^{M}}\;\frac{\overrightarrow{\partial}}{\partial
X^{^{M}}} B_{_{NP}}+(-1)^{^{N+M(N+P)}}\;\frac{\overrightarrow{\partial}}{\partial
X^{^{N}}} B_{_{PM}}+(-1)^{^{P(1+M+N)}}\;\frac{\overrightarrow{\partial}}{\partial
X^{^{P}}} B_{_{MN}}.
\end{eqnarray}
In Poisson-Lie T-duality, the duality transformation must include a correction term originating from the path-integral
integration of the fields on the dual group \cite{vonUnge}. This term is absorbed into the dilaton transformation at the one-loop level.
Following this prescription, the super Poisson-Lie T-duality transformations are given by\footnote{ ``sdet'' stands for superdeterminant \cite{D}.}
\begin{eqnarray}
\Phi \hspace{-1.1mm} &=& \hspace{-1.1mm} \phi^{^{(0)}}\hspace{-1.4mm} +\frac{1}{2}\log \big|s\hspace{-0.6mm}\det\big({_{_a}}{E_{_b}}(g, y^i)\big) \big| -\frac{1}{2}\log \big|s\hspace{-0.6mm}\det\big({_{_a}}{{E_0}_{_b}}(y^i)\big)\big| -\frac{1}{2}\log \big|s\hspace{-0.6mm}\det\big({_{_a}}{a^{^{b}}} { (g)}\big)\big|,~~~\label{2.17}\\
{\tilde \Phi} \hspace{-1.1mm} &=& \hspace{-1.1mm} \phi^{^{(0)}} \hspace{-1.4mm} +\frac{1}{2} \log |s\hspace{-0.6mm}\det\big({{{\tilde E}^{ab}}}(\tilde g, y^i)\big)| - \frac{1}{2}
\log|s\hspace{-0.6mm}\det\big({{{\tilde a}^a}_{_{~}b} ({\tilde g})}\big)|,\label{2.18}
\end{eqnarray}
where $\phi^{^{(0)}}$ is the dilaton that makes the original $\sigma$-model conformal up to the one-loop order
and may depend on both supergroup and spectator coordinates.

\section{Super non-Abelian T-duality of the Schwarzschild metric coupled to two fermionic fields}
\label{Sec.III}

Our goal in this section is to derive non-Abelian target space duals of the Schwarzschild spacetime, which is deformed by two fermionic fields.
We perform this analysis using the super Poisson-Lie T-duality on a semi-Abelian superdouble.
The Lie supergroup acting freely on the original target supermanifold of $T$-dual $\sigma$-models is considered to be one of the $(2|2)$-dimensional
non-Abelian Lie supergroups $({C}^3 +{A})$, ${C}^3 \otimes {A}_{1,1}$, and $(2{A}_{1,1}+2{A})^0$
whose Lie superalgebras obey the following set of (anti-)commutation relations \cite{B,ER6}:
\begin{align}
({\C}^3 +{\A}):~~~~&{[T_{_1} , T_{_4}]} = T_{_3},~~~~~\{T_{_4} , T_{_4}\}=T_{_2},\label{3.1}\\
{\C}^3 \oplus {\A}_{1,1}:~~~~&{[T_{_1} , T_{_4}]} = T_{_3},\label{3.2}\\
(2{ \A}_{1,1}+2{ \A})^0:~~~~&\{T_{_3} , T_{_3}\}=T_{_1}.\label{3.3}
\end{align}
The above superalgebras possess two bosonic generators $(T_{_1}, T_{_2})$ along with two
fermionic ones. The latter are denotes by $(T_{_3}, T_{_4})$.
Note that for investigating the super non-Abelian T-duality, the Lie supergroup of dual target supermanifold must be chosen to be Abelian.
In fact, we are dealing with the semi-Abelian superdoubles $(({\C}^3 + {\A})\;,\;{\cal I}_{_{(2|2)}})$,
$({\C}^3 \oplus {\A}_{1,1}\;,\;{\cal I}_{_{(2|2)}})$ and $((2{\A}_{1,1}+2{ \A})^0 , {\cal I}_{_{(2|2)}})$,
which are non-isomorphic as Lie superalgebras in each of the models.
Here we have denoted Abelian Lie superalgebra of the type $(2|2)$ by ${\cal I}_{_{(2|2)}}$.


\subsection{Super non-Abelian T-duality starting from the superdouble\\ $(({\C}^3 + {\A})\;,\;{\cal I}_{_{(2|2)}})$}
\label{III.1}

In what follows we shall obtain a conformal supersymmetric deformation of the Schwarzschild spacetime from a $T$-dualizable $\sigma$-model
constructing on a six-dimensional target supermanifold ${\M} \approx O \times  G$ where $ G$ as the first sub-supergroup of
Drinfeld superdouble is considered to be the $(C^{3} +A)$
acting freely on ${\M}$, while $O$ as the orbit of ${ G}$ in ${\M}$ is a two-dimensional space with the coordinates $y^i = (r,\theta)$.
The second sub-supergroup, ${\tilde { G}} = {I}_{_{(2|2)}}$, acting freely
on the dual supermanifold ${\tilde {\M}} \approx O \times {\tilde { G}}$
is assumed to be Abelian of the type $(2|2)$, and hence, the super Poisson-Lie T-duality reduces to the super non-Abelian T-duality.
Since the super Poisson-Lie duality is based on the concepts of the Drinfeld superdouble, it is necessary to form the semie-Abelian superdouble
$(({\C}^3 + {\A})\;,\;{\cal I}_{_{(2|2)}})$. Using \eqref{2.3} and \eqref{3.1}, it is easily deduced that
the Lie superalgebra of superdouble $(({\C}^3 + {\A})\;,\;{\cal I}_{_{(2|2)}})$ obeys the following non-vanishing (anti-)commutation relations
\begin{eqnarray}\label{3.4}
{[T_{_1} , T_{_4}]} &=& T_{_3},~~~~~~~~~~\{T_{_4} , T_{_4}\}=T_{_2},~~~~~~~\{T_{_1} , {\tilde T}^3\}=-{\tilde T}^4,\nonumber\\
{[T_{_4} , {\tilde T}^2]} &=& -{\tilde T}^4,~~~~~~~\{T_{_4} , {\tilde T}^3\}=-{\tilde T}^1,
\end{eqnarray}
where $\{T_{_a}\}$  and $\{{\tilde T}^a\}$ generate the $({\C}^3+{\A})$ and ${\cal I}_{_{(2|2)}}$, respectively.
In order to calculate the right-invariant super one-forms
one might suggest the following expansion for a generic element of the $(C^3 +A)$
\begin{eqnarray}\label{3.5}
g~=~ e^{\psi T_{_3}} ~ e^{x_{_1} T_{_1}}~e^{x_{_2} T_{_2}} ~e^{\chi T_{_4}},
\end{eqnarray}
where $(x_{_1}, x_{_2})$ are bosonic fields, while $(\psi, \chi)$ denote fermionic fields (Grassmann variables).
From now on, we will use this notation for bosonic and fermionic fields.
The right-invariant super one-forms are then read off to be
\begin{eqnarray}\label{3.6}
R_{\pm}^1&=& \partial_{\pm} x_{_{1}},~~~~~~~~~~~~~~~~~~~~~~~~~R_{\pm}^2 ~=~ \partial_{\pm} x_{_{2}}
+ \partial_{\pm} \chi ~ \frac{\chi}{2},\nonumber\\
R_{\pm}^3&=& -\partial_{\pm} \psi- \partial_{\pm} \chi~  x_{_{1}}, ~~~~~~~~~~R_{\pm}^4 ~=~ -\partial_{\pm} \chi.
\end{eqnarray}
Furthermore, using \eqref{3.4} and also equations \eqref{2.11} and \eqref{3.5} and their dual versions we can construct the super Poisson structures
on both Lie supergroups $(C^3 +A)$ and ${I}_{_{(2|2)}}$, giving
\begin{eqnarray}
{\Pi}^{ab} =0,~~~~~~~~~~~~ {\tilde \Pi}_{ab}=\left( \begin{tabular}{cccc}
                 0 & 0 & 0 & -$ {\tilde \psi}$ \\
                 0 & $0$  & 0& 0 \\
                 0 & 0& $0$ & 0 \\
                 $ {\tilde \psi}$ & 0 & 0 & -${\tilde x}_{_2}$ \\
                 \end{tabular} \right),\label{3.7}
\end{eqnarray}
where we have chosen $({\tilde x}_{_1}, {\tilde x}_{_2}; {\tilde \psi}, {\tilde \chi})$ as the supercoordinates of the dual supergroup.
In order to write the original $\sigma$-model on the ${\M}$ with the $(C^3 +A)$ Lie supergroup in the form of the action \eqref{2.8}
we need to determine the model couplings.
Let us choose the spectator-dependent matrices in the following form
{\small \begin{eqnarray}
{E_{0}}_{_{ab}}=\left( \begin{tabular}{cccc}
                 $-(1-\frac{2M}{r})$ & 0 & 0 & 0 \\
                 0 & $r^{2} \sin^2 \theta$  & 0& 0 \\
                 0 & 0& $n (r)$ & $F(r)$ \\
                 0 & 0 & -$F(r)$ & $h(r)$ \\
                 \end{tabular} \right),~ F_{i j} = \left( \begin{tabular}{cccc}
                 $(1-\frac{2M}{r})^{-1}$ & 0 \\
                 0 & $r^{2}$ \\
                 \end{tabular} \right),~F^{(1)}_{a i}  = F^{(2)}_{i b} = 0.\label{3.8}
\end{eqnarray}}
where the parameter $M$ is the mass of the object creating the gravitational field in metric \eqref{1.1}, and
$n(r)$, $h(r)$ and $F(r)$ are arbitrary functions of $r$, such that $F(r)$ cannot be zero.
According to \eqref{3.7} the super Poisson structure on the $(C^3 +A)$ is zero.
On the other hand, as mentioned in Sec. \ref{Sec.II}, in the super non-Abelian T-duality case,
the various couplings in the action \eqref{2.8} turn to
${{E}} = E_{0},~ \phi^{{(1)}} = F^{^{(1)}},~ \phi^{{(2)}} = F^{^{(2)}}$ and $\phi = F $.
Thus, by substituting \eqref{3.8} into \eqref{2.10} and then by applying \eqref{2.8}, one can obtain the action of the original model.
The corresponding line element and $B$-field are then read off
\begin{align}
ds^2=& (1-\frac{2M}{r})^{-1} ~ dr^2  -(1-\frac{2M}{r}) ~ d t^2  + r^{2} ({d \theta}^2+\sin^2 \theta ~ d \varphi^2) \nonumber\\
&~~~~~~~~~~~~~~~~~~~~~~~~~~~~~~~~~~~~~~~~~~~~ - r^{2} \chi \sin^2 \theta  ~ d \varphi d \chi-2 F(r) ~ d\psi d\chi,\label{3.9}\\
B=& -\frac{1}{2} n(r)~ d\psi \wedge d\psi- t ~n (r)~ d\psi \wedge d\chi-\frac{1}{2} \big(t^2 n(r)+h(r)\big)~ d\chi \wedge d\chi.\label{3.10}
\end{align}
Here we have used the change of coordinates $x_{1} \rightarrow t$ and $x_{2}\rightarrow \varphi$.
As can be seen from \eqref{3.9}, the bosonic part is exactly the Schwarzschild geometry, metric \eqref{1.1}, and
the fermionic contribution (deformation) enters through $- r^{2} \chi \sin^2 \theta  ~ d \varphi d \chi$ and $-2 F(r) ~ d\psi d\chi$.
These terms are linear in Grassmann variables.
Because Grassmann variables are nilpotent, $\psi^2 =\chi^2=0$, all quadratic contractions involving fermionic corrections vanish.
Hence, the Ricci scalar will not receive any non-trivial nilpotent corrections.
Thus, the supersymmetric deformation preserves the scalar-flat property of Schwarzschild spacetime.

Below we want to determine the functions $n(r)$, $h(r)$ and $F(r)$ by imposing the one-loop conformal invariance conditions.
Before doing this, one may apply formula \eqref{2.17} to calculate dilaton field that makes the model conformal up to the one-loop order.
At first, by using the first equation of \eqref{2.11} and (anti-)commutation relations in \eqref{3.1}
one immediately concludes that sdet$\big({_{_a}}{a^{{b}}} { (g)}\big)=1$. Hence, from  \eqref{2.17} we find that $\Phi =\phi^{^{(0)}}$.
Looking at the equations \eqref{2.13}-\eqref{2.15} we conclude that these equations are
satisfied with the metric \eqref{3.9} and $B$-field \eqref{3.10} together with $\phi^{^{(0)}} = {\cal C}_{_0} $ if one considers
\begin{eqnarray}
n (r)= 0,~~~~h(r)= C_{_1}+\frac{C_{_2}}{2M}~\log|1-\frac{2M}{r}|,~~~~~F(r) =F_{_0},\label{3.11}
\end{eqnarray}
where $C_{_1}$, $C_{_2}$ and $F_{_0}$ are some arbitrary constants. In this way, the cosmological constant becomes zero.
As shown in \cite{de.laOssa},
the Schwarzschild metric with a constant $B$-field (vanishing field strength) satisfies the beta-function equations
in the absence of fermions. Upon coupling two fermions to the metric, the $B$-field ceases to be constant,
and the beta-function equations are satisfied with a field strength of the form  $H= \frac{C_{_2}}{ r(r-2M)}~ dr \wedge d\chi \wedge d\chi$.
Consequently, we have constructed a conformal supersymmetric deformation of Schwarzschild spacetime by the super non-Abelian T-duality.

Taking into account \eqref{3.11}, the metric \eqref{3.9} appears to have singularities at $r = 0$ and $r=2M$.
The case $r = 0$ is different, however. If one asks that the solution be valid for all $r$
one runs into a true physical singularity at the origin. To see that this is a true singularity one must look at quantities that are independent of the choice of coordinates.
One such important quantity is the Kretschmann invariant, being the square of the Riemann
tensor, i.e. ${\cal K} = {\cal R}_{_{MNPQ}} {\cal R}^{^{QPNM}}$.
The other quantities are the Ricci scalar ${\cal R}$ and ${\cal R}_{_{MN}} {\cal R}^{^{NM}}$.
For the above metric we find that ${\cal R}$ and ${\cal R}_{_{MN}}$ vanish, whereas the Kretschmann invariant is
\begin{eqnarray}
{\cal K}= \frac{48M^{2}}{ r^{6}}.\label{3.12}
\end{eqnarray}
Thus, the fermionic extension does not introduce any additional curvature singularities into the Kretschmann invariant.
It is more appropriate to say that the nilpotent fermionic terms do not modify the classical Schwarzschild
curvature invariants.
The only true curvature singularity occurs at $r=0$, where ${\cal K} \propto { r^{-6}}$ diverges. In contrast,
at $r=2M$ the Kretschmann invariant remains finite, ${\cal K}(2M)=3/(4M^4)$, although $G_{rr}$
diverges in Schwarzschild coordinates. Moreover, since $G^{rr}=1-2M/r$ vanishes at $r=2M$,
this hypersurface is null and, being non-singular in curvature invariants, represents a regular event horizon rather than a curvature singularity.
Consequently, the central singularity at $r=0 $ is hidden behind the regular horizon $r=2M$, and the geometry does not contain
a naked curvature singularity.

In order to add some interpretation of the background, it is useful to look at the superisometry symmetries of the metric.
To this end, one must calculate the corresponding Killing supervectors.
In this way, we use the graded form of Killing equation \cite{samadi,Meysam2}
\begin{align}\label{3.13}
{\cal L}_{_{K_a}} {G}_{_{MN}} = (-1)^{M +P +M a} \frac{\overrightarrow{\partial} {K_a}^{P}}{{\partial} X^{^{M}}} {G}_{P N}&+{K_a}^{P} \frac{\overrightarrow{\partial} {G}_{MN}}{{\partial} X^{^{P}}}\nonumber\\
&+ (-1)^{MN +M P +P+ N a +N} \frac{\overrightarrow{\partial} {K_a}^{P}}{{\partial} X^{^{N}}} {G}_{M P} = 0,
\end{align}
where the Killing supervector $K_a$ is considered to have two degrees, even (bosonic Killings indicated by $K^B_{_a}$) and odd (fermionic
Killings indicated by $K^F_{_a}$).
Using the above formula, we can analyze the superisometry structure of the geometry.
The metric \eqref{3.9} admits a set of three bosonic Killing vectors,
denoted as $K^B_{_a}$, which generate the isometry group of the spacetime. These are given by
\begin{eqnarray}\label{3.14}
K^B_{_1} = \overrightarrow{\frac{\partial}{\partial t}},~~~~~~~~
K^B_{_2} = \overrightarrow{\frac{\partial}{\partial \varphi}},~~~~~~~
K^B_{_3} = \chi \overrightarrow{\frac{\partial}{\partial \psi}}.
\end{eqnarray}
It can be easily shown that the algebra generated by these three vectors is Abelian,
while the isometry group of the purely bosonic Schwarzschild metric (metric \eqref{1.1}) is $SO(3) \times U(1)$.
This result shows that the supersymmetric deformation, unlike the Schwarzschild
curvature invariants, has affected the isometry symmetries of the metric.
The norm analysis reveals a clear separation between the ordinary bosonic causal structure
and the additional fermionic symmetries of the supermanifold.
The causal character of these Killing vectors is determined from their norms,
$|| K_{_a}||^2 = (-1)^{^M} K^{^M}_{_a} G_{_{MN}} K^{^N}_{_a}$.
For the exterior region $r > 2M$, the Killing vector $K^B_{_1}$ with the norm $||K^B_{_1}||^2 = -1+\frac{2M}{r}$ is a negative time vector.
This allows stationary observers in this region to evolve in time.
Inside the horizon, $r < 2M$, the term $-1+\frac{2M}{r}$ becomes positive.
Thus, for the interior region, the $K^B_{_1}$ is a spacelike vector, and it becomes null on the event horizon $(r=2M)$.
The norm of the vector $K^B_{_2}$ is $r^2 \sin^2 \theta$. In almost all regions of the spacetime, specifically where $r>0$ and
$\theta \neq 0, \pi$, the norm is positive. Therefore, in the vast majority of the spacetime, the $K^B_{_2}$ is a spacelike vector.
This aligns with our understanding that this vector represents a circular displacement (rotation) in three-dimensional space.
Moving along this direction constitutes a spatial displacement rather than a temporal evolution. Also, this
vector becomes null when its norm vanishes. This occurs in two specific geometric scenarios:
(1) on the axis of rotation $(\theta = 0$ or $\theta = \pi)$, at the poles of the coordinate system, $\sin \theta =0$.
(2) at the central singularity $(r=0)$. It is worth stressing that this vector is never timelike.
For the Killing vector $K^B_{_3}$ we find that its norm is zero and thus becomes null.

Furthermore, the geometry exhibits supersymmetry, characterized by two fermionic Killing vectors, $K^F_{_1}$ and $K^F_{_2}$, defined as:
\begin{eqnarray}\label{3.15}
K^F_{_1} =  \overrightarrow{\frac{\partial}{\partial \psi}},~~~~~~~~~~~~
K^F_{_2} =  \frac{\chi}{2} \overrightarrow{\frac{\partial}{\partial \varphi}} + \overrightarrow{\frac{\partial}{\partial \chi}}.
\end{eqnarray}
The fermionic Killing vectors $K^F_{_1}$ and $K^F_{_2}$ are found to be null, as their associated norm vanishes identically across the
target supermanifold.
Note that all fermionic Killing vectors have vanishing self-contraction with respect to the even supermetric.
This is a natural consequence of the graded symmetry of the supermetric for Grassmann-odd vector fields
and should not be confused with the usual notion of null directions in a Lorentzian bosonic manifold.
The usual causal structure is defined in terms of the commuting bosonic coordinates, whereas the Grassmann-odd coordinates describe additional supersymmetric degrees of freedom.
In this sense, $K^F_{_1}$ and $K^F_{_2}$ are better regarded as generators of fermionic symmetries of the supermanifold
rather than as tangent directions corresponding to the worldlines of ordinary physical observers.
To uncover the underlying algebraic structure, we calculate the (anti-)commutation relations associated with these generators.
The Lie superalgebra is spanned by the bosonic and fermionic generators, satisfying the following non-vanishing independent relations:
\begin{eqnarray}\label{3.16}
[K^B_{_3}  , K^F_{_2}] =  -K^F_{_1},~~~~~~~~~\{K^F_{_2}  , K^F_{_2}\} =  K^B_{_2}.
\end{eqnarray}
These relations characterize the non-trivial structure constants of the Killing superalgebra associated with the metric.
It is straightforward to verify that the Lie superalgebra spanned by these vectors is isomorphic to
$({\C}^3 +{\A}) \oplus {\A}_{1,1}$, where $({\C}^3 +{\A})$ is defined by \eqref{3.1} and ${\A}_{1,1}$
denotes the one-dimensional Abelian Lie algebra generated by a single bosonic generator $K^B_{_1}$.

To continue, we can obtain a dual pair for background given by equations \eqref{3.9} and \eqref{3.10}.
The dual model is constructed on the supermanifold ${\tilde {\M}} \approx O \times {\tilde { G}}$, where ${\tilde { G}}$
is the Abelian Lie supergroup ${I}_{_{(2|2)}}$ which is parameterized with the supercoordinates $({\tilde x}_{_1}, {\tilde x}_{_2}; {\tilde \psi}, {\tilde \chi})$, as mentioned a little earlier, and
$O$ will be the orbit introduced in the supermanifold ${\M}$ with the same coordinates $y^i = (r,\theta)$.
Now, substituting \eqref{3.7} and \eqref{3.8} into \eqref{2.12}, one can get the dual couplings. They are then read
\begin{eqnarray}
{\tilde E}^{^{ab}} =\left( \begin{tabular}{cccc}
                 -$\frac{1}{1-\frac{2M}{r}}$ & 0 & $\frac{{\tilde \psi}}{F_{0}(1-\frac{2M}{r})}$ & 0 \\
                 0 & $\frac{1}{r^{2} \sin^2 \theta}$  & 0 & 0 \\
                 -$\frac{{\tilde \psi}}{F_{0}(1-\frac{2M}{r})}$  & 0 & $ \frac{{\tilde x}_{_{2}} -h(r)}{F_{0}^{2}}$ & $\frac{1}{F_{0}}$ \\
                 0 & 0 & -$\frac{1}{F_{0}}$ & 0 \\
                 \end{tabular} \right),~~~ {\tilde \phi}^{\hspace{0mm}{(1)^{ a}}}_{~~~j} ={\tilde \phi}^{\hspace{0mm}{(2)^{ b}}}_{i}=0,~~~{\tilde \phi}_{_{ij}}=F_{_{ij}},\label{3.17}
\end{eqnarray}
where $F_{_{ij}}$ and $h(r)$ are defined by equations \eqref{3.8} and \eqref{3.11}, respectively.
Substituting \eqref{3.17} into the action \eqref{2.9} and then using the fact that the components of the right-invariant super one-forms on
${I}_{_{(2|2)}}$ are ${\tilde R}_{\pm_{a}}=\partial_{\pm} {{\tilde x}_a}$, one can write down the action of dual $\sigma$-model.
The corresponding line element and $\tilde B$-field are worked out to be
\begin{align}
{\tilde ds^2}=& \frac{1}{1-\frac{2M}{r}}  dr^2 + r^{2} ~ {d \theta}^2 -\frac{1}{1-\frac{2M}{r}}  {d {\tilde x}_{_{1}}}^2  + \frac{1}{r^{2} \sin^2\theta} {d {\tilde x}_{_{2}}}^2 - \frac{2}{F_{0}} ~ d {\tilde \psi} d{\tilde \chi},\label{3.18}\\
{\tilde B}=& -\frac{{\tilde \psi}}{F_{0}(1-\frac{2M}{r})} ~  {d {\tilde x}_{_{1}}} \wedge d{\tilde \psi} -\frac{{\tilde x}_{_{2}}-h(r)}{2{F_{0}}^{2}}  ~ d {\tilde \psi} \wedge d {\tilde \psi}.\label{3.19}
\end{align}
Using the \eqref{2.16}, the field strength corresponding to the $\tilde B$-field is computed to be
\begin{align}
\tilde{H}=& \frac{2M \tilde{\psi}}{ F_{0}(r-2M)^{2}}~ dr \wedge d\tilde{x_{1}} \wedge d\tilde{\psi} - \frac{C_{2}}{r {F_{0}}^{2}(r-2M)}~ dr \wedge d\tilde{\psi} \wedge d\tilde{\psi} \nonumber\\
  &~~~~~~~~~~~~~~~~~~~- \frac{2r}{{F_{0}}(r-2M)}~ d\tilde{x_{1}} \wedge d\tilde{\psi} \wedge d\tilde{\psi} + \frac{1}{{F_{0}^{2}}}~ d\tilde{x_{2}} \wedge d\tilde{\psi} \wedge d\tilde{\psi}. \label{3.20}
\end{align}
In order to investigate the conformal invariance conditions of the dual background, let us evaluate the dual dilaton.
Since the dual Lie supergroup is Abelian, using the dual version of the first equation of \eqref{2.11}, one can get
${{\tilde a}^{^a}}_{~b}({\tilde g}) = {{\delta}^{^a}}_{b}$; consequently, $s\hspace{-0.6mm}\det({{\tilde a}^{^a}}_{~b}({\tilde g})) =1 $.
Also we find that $s\hspace{-0.6mm}\det({\tilde E}^{{ab}}) = {{F_{0}}^{2} r^{-2} (\frac{2M}{r} -1)^{-1}} \csc^2 \theta $ and hence
we get the dual dilaton from equation \eqref{2.18} by remembering that $\phi^{^{(0)}}= {\cal C}_{_0} $ which gives the final result
\begin{eqnarray}\label{3.21}
{\tilde \Phi} = {\cal C}_{_0} +\frac{1}{2}\log \Big|\frac{{F_{0}}^{2}}{r^{2} \sin^2 \theta ~ (\frac{2M}{r}-1)} \Big|.
\end{eqnarray}
Now, one can check that the field equations \eqref{2.13}-\eqref{2.15} are satisfied for the dual metric
\eqref{3.18}, field strength \eqref{3.20} and dilaton field \eqref{3.21} together with a zero cosmological constant.
This means that the dual background is also conformally invariant up to one-loop order.

On the other hand, we are interested in investigating the structure of dual geometry.
For the dual geometry described by the above metric, the curvature invariants reveal a structure that is qualitatively different
from the original spacetime. We first find that the scalar curvature of the dual metric is
\begin{eqnarray}
\tilde{{\cal R}} = \frac{2 \big(4Mr-M^{2}-2r^{2}+M^{2}\cos2\theta \big)}{(r-2M)r^{3} \sin^{2}\theta}. \label{3.22}
\end{eqnarray}
Notably, the Kretschmann invariant behaves as
\begin{eqnarray}
\tilde{{\cal K}} \propto \frac{1}{ r^{6}(r-2M)^{2} \sin^{4} \theta}, \label{3.23}
\end{eqnarray}
and therefore diverges at $r=0$, $r=2M$, and $\theta=0, \pi$.
Hence, these loci correspond to true curvature singularities rather than mere coordinate artifacts.
In particular, although $G^{rr}=1-2M/r$ vanishes at $r=2M$, indicating that this hypersurface is null,
the divergence of $\tilde{{\cal K}}$ and $\tilde{{\cal R}}$ demonstrates that $r=2M$ is itself a null curvature singularity and cannot be interpreted
as a regular Schwarzschild-like event horizon.
Thus, unlike the Schwarzschild case, the would-be horizon at $r=2M$ is not a removable coordinate singularity and the usual regular horizon structure is absent. The loci $\theta=0, \pi$ are also true curvature singularities, as confirmed by the
$1/\sin^{4} \theta$ divergence, while $r=0$ exhibits the characteristic strong divergence ${\tilde{{\cal K}}} \propto r^{-6}$.
Consequently, the geometry contains singular null surfaces rather than a regular black-hole horizon, making $r=2M$ a natural candidate for a naked null singularity, subject to confirmation through a global causal-structure analysis.
In general relativity, a naked singularity is a hypothetical gravitational singularity without an event horizon.
The theoretical existence of naked singularities is important because their existence would mean that it would be possible to observe the collapse of an object to infinite density.

Having identified the singularities, we now turn our attention to the symmetries of the dual metric.
The dual metric exhibits an enhanced symmetry structure compared to the original geometry.
Upon introducing the coordinate transformation  ${\tilde x}_{_{1}} \rightarrow t$ and $\rightarrow {\tilde x}_{_{2}} \rightarrow \varphi$,
the metric exhibits five bosonic Killing vectors, ${\tilde K}^B_{_a} (a=1, ..., 5)$, and two fermionic Killing vectors, ${\tilde K}^F_{_a} (a=1,2)$, which are given by
\begin{align}\label{3.24}
{\tilde K}^B_{_1} =& \overrightarrow{\frac{\partial}{\partial t}},~~~~~~~
{\tilde K}^B_{_2} = \overrightarrow{\frac{\partial}{\partial \varphi}},~~~~~~
{\tilde K}^B_{_3} = \psi \overrightarrow{\frac{\partial}{\partial \psi}}-\chi \overrightarrow{\frac{\partial}{\partial \chi}},~~~~~
{\tilde K}^B_{_4} = \chi \overrightarrow{\frac{\partial}{\partial \psi}},~~~~~
{\tilde K}^B_{_5} =\psi \overrightarrow{\frac{\partial}{\partial \chi}},
\nonumber\\
{\tilde K}^F_{_1} =&  \overrightarrow{\frac{\partial}{\partial \psi}},~~~~~~
{\tilde K}^F_{_2} =   \overrightarrow{\frac{\partial}{\partial \chi}}.
\end{align}
The Killing vector ${\tilde K}^B_{_1}$ is the unique generator among the seven listed Killing supervectors that
provides a conventional timelike direction in the bosonic part.
Its norm is negative for $r> 2M$, thereby defining the stationary timelike region of the geometry.
This result confirms that the T-duality has been involved the timelike directions.
The hypersurface $ r=2M $ is distinguished by the change in the causal character of the stationary Killing vector. Outside this surface,
the time-translation generator is timelike, whereas inside it becomes spacelike.
We find that the norm of second bosonic Killing vector is $||{\tilde K}^B_{_2}||^2={1}/({r^2\sin^2\theta})$.
Indeed, this quantity is positive over the regular coordinate domain $ r>0$, $ 0<\theta<\pi$. Therefore, ${\tilde K}^B_{_2}$ becomes spacelike.
This is consistent with its interpretation as the generator of rotations in the azimuthal direction.
The norm of ${\tilde K}^B_{_3}$, on the other hand, is proportional to the Grassmann-even bilinear $\psi \chi$, i.e., $||{\tilde K}^B_3||^2=2\psi \chi/F_0$,
up to the convention adopted for the ordering of the Grassmann variables.
This is fundamentally different from the norm of an ordinary bosonic spacetime vector.
The important point is that $\psi \chi$ is Grassmann-even but nilpotent.
Consequently, this norm is not an ordinary real-valued scalar and cannot be classified as positive or negative in the usual Lorentzian sense.
Therefore, it is not appropriate to characterize ${\tilde K}^B_{_3}$ as timelike or spacelike by means of the conventional sign criterion.
The remaining two bosonic Killing vectors are ${\tilde K}^B_{_4}$ and ${\tilde K}^B_{_5}$.
Because the fermionic contribution to the metric is purely off-diagonal, namely,
${\tilde G}_{\psi \psi}={\tilde G}_{\chi \chi}=0$,
the norms of ${\tilde K}^B_4$ and \({\tilde K}^B_5\) vanish identically.
These vectors are null in the supergeometric sense. Nevertheless, their vanishing norms should not be interpreted as ordinary null propagation along
bosonic spacetime light cones. Both vectors are associated entirely with the Grassmann part of the supermanifold.
In particular, neither ${\tilde K}^B_{_4}$ nor ${\tilde K}^B_{_5}$ provides an additional physical time direction.
The fermionic Killing vectors ${\tilde K}^F_{_1}$ and ${\tilde K}^F_{_2}$ also have vanishing self-norms, although they have a non-zero mutual inner product.
Therefore, the vanishing norms of the fermionic Killing vectors should be understood as a characteristic feature of the graded geometry rather than as evidence for additional physical null directions.

In order to clarify the underlying algebraic structure, we compute the (anti-)commutation relations of these generators.
The resulting Lie superalgebra, spanned by the seven bosonic and fermionic generators mentioned above, satisfies the following non-vanishing relations:
\begin{align}\label{3.25}
[{\tilde K}^B_{_3}  , {\tilde K}^B_{_4}] =&  -2 {\tilde K}^B_{_4},~~~~~~~[{\tilde K}^B_{_3}  , {\tilde K}^B_{_5}] = 2 {\tilde K}^B_{_5},~~~~~~~[{\tilde K}^B_{_5}  , {\tilde K}^B_{_4}] =  {\tilde K}^B_{_3},~~~~~~
[{\tilde K}^B_{_3} , {\tilde K}^F_{_1}] = -{\tilde K}^F_{_1}\nonumber\\
[{\tilde K}^B_{_3} , {\tilde K}^F_{_2}] =& {\tilde K}^F_{_2},~~~~~~~~~~~~[{\tilde K}^B_{_4} , {\tilde K}^F_{_2}] = -{\tilde K}^F_{_1},~~~~~~~[{\tilde K}^B_{_5} , {\tilde K}^F_{_1}] = -{\tilde K}^F_{_2}.
\end{align}
It is evident that the bosonic subalgebra spanned by four generators $\{{\tilde K}^B_{_3}, {\tilde K}^B_{_4}, {\tilde K}^B_{_5}\}$ is isomorphic to
$sl (2 , \Re)$. Furthermore, ${\tilde K}^B_{_1}$ and ${\tilde K}^B_{_2}$ are central generators, as they commute with all other generators in the algebra.

In summary, within this subsection, we have starting from the superdouble $(({\C}^3 +{\A}) , {\cal I}_{_{(2|2)}})$,
demonstrated a concrete example of super Poisson-Lie T-duality. This process has yielded a non-Abelian target space dual (described by equations \eqref{3.18} and \eqref{3.19})
for the Schwarzschild spacetime deformed by two Grassmann fields.


\subsection{Super non-Abelian T-duality starting from the superdouble \\$({\C}^3 \oplus {\A}_{1,1}\;,\;{\cal I}_{_{(2|2)}})$}
\label{III.2}

In this subsection, we construct an additional supersymmetric deformation of the Schwarzschild geometry within the
super Poisson-Lie duality framework. This yields a new non-Abelian dual pair, whose distinction from the preceding dual background resides solely in the corresponding ${\tilde B}$-field.
The Lie supergroup associated with the target space supermanifold  ${{\M}} \approx O \times  G$ is identified as
$C^3 \otimes { A}_{1,1}$ whose Lie superalgebra is defined by \eqref{3.2}.
The orbit $O$ of $G$ is defined in a manner analogous to the previous example, employing the same coordinates $y^i = (r,\theta)$.
The Lie superalgebra of semie-Abelian Drinfeld superdouble
$({\C}^3 \oplus {\A}_{1,1}\;,\;{\cal I}_{_{(2|2)}})$ is determined by the following (anti-)commutation relations:
\begin{eqnarray}\label{3.26}
{[T_{_1} , T_{_4}]} = T_{_3},~~~~~~~~~~{[T_{_1} , {\tilde T}^3]} = -{\tilde T}^4,~~~~~~~~\{T_{_4} , {\tilde T}^3\}=-{\tilde T}^1.
\end{eqnarray}
For the parametrization of a general element of ${C}^3 \otimes {A}_{1,1}$ with the supercoordinates $(x_{_{1}}, x_{_{2}}; \psi, \chi)$,
adopting the choice used in \eqref{3.5}, we obtain
\begin{eqnarray}\label{3.27}
R_{\pm}^1&=& \partial_{\pm} x_{_{1}},~~~~~~~~~~~~~~~~~~~~~~~~R_{\pm}^2 ~=~ \partial_{\pm} x_{_{2}},\nonumber\\
R_{\pm}^3&=& -\partial_{\pm} \psi- \partial_{\pm} \chi~  x_{_{1}}, ~~~~~~~~~R_{\pm}^4 ~=~ -\partial_{\pm} \chi.
\end{eqnarray}
In order to construct the original $\sigma$-model on the $C^3 \otimes { A}_{1,1}$,
we adopt the spectator-dependent matrices following the form of \eqref{3.8}, where the functions
$n(r), h(r),$ and $F(r)$ are determined by \eqref{3.11}.
Furthermore, since the dual Lie supergroup is identified as Abelian,
it follows from \eqref{2.11} that the super Poisson structure on the $C^3 \otimes { A}_{1,1}$ vanishes.
Combining these results, the action of the model on the $C^3 \otimes { A}_{1,1}$ is obtained.
The corresponding background, including the line element and the $B$-field, is given by
\begin{align}
ds^2 =& \frac{1}{1-\frac{2M}{r}} ~ dr^2 - (1-\frac{2M}{r}) ~ {d t}^2 + r^{2} ({d \theta}^2 +\sin^2 \theta ~ {d \varphi}^2)
-2 F_0 ~ d\psi d\chi,\label{3.28} \\
B =& -\frac{1}{2} h(r)~ d\chi \wedge d\chi.\label{3.29}
\end{align}
Following our previous convention, we have applied the change of variables $x_{1} \rightarrow t$ and $x_{2}\rightarrow \varphi$.
Interestingly, the resulting background preserves the standard Schwarzschild metric in its bosonic part,
leaving the angular part free from Grassmann-dependent deformations.
Its fermionic contribution is restricted to the bilinear term $-2 F_0 ~ d\psi d\chi$.
The metric exhibits a singularity structure and behavior analogous to the metric presented in \eqref{3.9}.
For the configuration \eqref{3.28}, the Ricci tensor and scalar curvature vanish identically,
and the Kretschmann invariant is consistent with that derived in \eqref{3.12}.
For the background defined by the metric and $B$-field, one may similarly determine the dilaton field required for conformal invariance up to the one-loop order.
The dilaton obtained from \eqref{2.17} reduces to a constant,
$\Phi = \phi^{^{(0)}} = {\cal C}_{_0}$.
Substitution of the metric and $B$-field into the one-loop beta-function equations shows that the equations are consistently satisfied for a constant dilaton field.
Thus, the new six-dimensional background remains conformal up to the one-loop order upon choosing the dilaton to be constant.
Both the $(C^{3} + A)$ and $C^3 \otimes { A}_{1,1}$ backgrounds demonstrate precisely this point:
they share identical bosonic limits, scalar curvatures, Kretschmann invariants, and functional forms of the $B$-field,
yet they differ in their Grassmann-dependent metric components and super Killing structures, as will be demonstrated below.
While the $C^3 \otimes { A}_{1,1}$ background exhibits the largest manifest superisometry sector,
the background from the previous example is considerably more constrained.
The $C^3 \otimes { A}_{1,1}$ background possesses seven bosonic Killing vectors,
\begin{align}\label{3.30}
K^B_{_1} =& \overrightarrow{\frac{\partial}{\partial t}},~~~~~~~
K^B_{_2} =\sin \varphi \overrightarrow{\frac{\partial}{\partial \theta}} +\cos \varphi \cot \theta \overrightarrow{\frac{\partial}{\partial \varphi}},~~~~~~
K^B_{_3} = \cos \varphi \overrightarrow{\frac{\partial}{\partial \theta}} -\sin \varphi \cot \theta \overrightarrow{\frac{\partial}{\partial \varphi}},\nonumber\\
K^B_{_4} =&  \overrightarrow{\frac{\partial}{\partial \varphi}},~~~~~~
K^B_{_5} =\psi \overrightarrow{\frac{\partial}{\partial \psi}} - \chi \overrightarrow{\frac{\partial}{\partial \chi}},~~~~~~~~~~~~~~~~~~~~~
K^B_{_6} = \chi \overrightarrow{\frac{\partial}{\partial \psi}},~~~~~~~~
K^B_{_7} = \psi \overrightarrow{\frac{\partial}{\partial \chi}},
\end{align}
as well as two fermionic Killing vectors,
\begin{align}\label{3.31}
K^F_{_1} =  \overrightarrow{\frac{\partial}{\partial \psi}},~~~~~~~~~~~~~~~
K^F_{_2} =   \overrightarrow{\frac{\partial}{\partial \chi}}.
\end{align}
The first four bosonic generators reproduce the expected time-translation and rotational symmetry $(SO(3) \times U(1))$ of the bosonic Schwarzschild spacetime.
The remaining three bosonic generators act entirely or predominantly in the fermionic directions and reveal an enlarged symmetry of the fermionic subspace.
In particular, $K^B_{_5}, K^B_{_6}$, and $K^B_{_7}$ generate non-trivial linear transformations among the Grassmann coordinates,
such that they constitute the $sl(2,\Re)$ Lie algebra.
The fact that both fermionic translations $\overrightarrow{\partial}/{\partial \psi}$ and $\overrightarrow{\partial}/{\partial \chi}$
are separately Killing vectors is another
indication that the fermionic part of this background is less geometrically constrained than that of the  $(C^{3} + A)$ background.

Let us now consider the structure of the dual $\sigma$-model.
In a manner similar to the dual $\sigma$-model construction discussed previously,
the Lie supergroup ${\tilde { G}}$ of the dual supermanifold  ${\tilde {\M}} \approx O \times {\tilde { G}}$ is taken to be Abelian.
In this case, the super Poisson structure on the dual Lie supergroup possesses only one non-vanishing component, namely
${\tilde \Pi}_{{{\tilde x}_{_{1}} {\tilde \chi}}} = -{\tilde \psi}$
which yield the following dual coupling matrices
\begin{eqnarray}
{\tilde E}^{{ab}} =\left( \begin{tabular}{cccc}
                 -$\frac{1}{1-\frac{2M}{r}}$ & 0 & $\frac{{\tilde \psi}}{F_{0}(1-\frac{2M}{r})}$ & 0 \\
                 0 & $\frac{1}{r^{2} \sin^2 \theta}$  & 0 & 0 \\
                 -$\frac{{\tilde \psi}}{F_{0}(1-\frac{2M}{r})}$  & 0 & $ \frac{h(r)}{F_{0}^{2}}$ & $\frac{1}{F_{0}}$ \\
                 0 & 0 & -$\frac{1}{F_{0}}$ & 0 \\
                 \end{tabular} \right),~~~~~{\tilde \phi}^{\hspace{0mm}{(1)^{ a}}}_{~~~j} ={\tilde \phi}^{\hspace{0mm}{(2)^{ b}}}_{i}=0,~~~{\tilde \phi}_{_{ij}} =F_{i j},\label{3.32}
\end{eqnarray}
where $F_{_{ij}}$ and $h(r)$ are given by equations \eqref{3.8} and \eqref{3.11}, respectively.
Finally, one can derive the dual $\sigma$-model, characterized by the following line element and $\tilde B$-field
\begin{align}
{\tilde ds^2} =& \frac{1}{1-\frac{2M}{r}}  dr^2 + r^{2} ~ {d \theta}^2 -\frac{1}{1-\frac{2M}{r}}  {d {\tilde x}_{_{1}}}^2  + \frac{1}{r^{2} \sin^2 \theta} {d {\tilde x}_{_{2}}}^2
- \frac{2}{F_{0}} ~ d {\tilde \psi} d{\tilde \chi},\label{3.33}\\
{\tilde B} =& -\frac{{\tilde \psi}}{F_{0}(1-\frac{2M}{r})} ~  {d {\tilde x}_{_{1}}} \wedge d{\tilde \psi} - \frac{h(r)}{{{2 F_{0}}^{2}}}  ~ d {\tilde \psi} \wedge d {\tilde \psi}.\label{3.34}
\end{align}
The dilaton field ensuring the conformality of the dual background, derived from \eqref{2.17}, mirrors the form of \eqref{3.21}. Verification shows that the metric \eqref{3.33},
${\tilde B}$-field \eqref{3.34}, and dilaton \eqref{3.21} satisfy the field equations \eqref{2.13}-\eqref{2.15}
with zero cosmological constant, confirming the one-loop conformal invariance of the dual background.
Because the metric \eqref{3.33} is identical to the dual metric of the $(C^{3} + A)$
background, its scalar curvature, Kretschmann invariant, and overall singularity structure are equivalent to those reported for the dual metric \eqref{3.18}.
A similar conclusion holds for the superisometries: the shared metric implies equivalent Killing supervector fields.
Consequently, while the two dual models are indistinguishable from a purely metric and isometric perspective, they are distinguished by their ${\tilde B}$-fields.


\subsection{Super non-Abelian T-duality starting from the superdouble \\$((2{\A}_{1,1}+2{ \A})^0 , {\cal I}_{_{(2|2)}})$}

In this subsection, demonstrating the flexibility of the super Poisson-Lie T-duality framework, we construct a further supersymmetric deformation of the Schwarzschild metric. This derivation proceeds analogously to our previous examples, with the distinction that the Lie supergroup $G$ of the supermanifold ${{\M}} \approx O \times  G$ is chosen as $(2{ A}_{1,1}+2{ A})^0$, with its Lie superalgebra governed by \eqref{3.3}.
Adopting the parametrization \eqref{3.5} with the supercoordinates $(x_{_{1}}=t, x_{_{2}}=\varphi; \psi, \chi)$ for the elements of the
$(2{ A}_{1,1}+2{ A})^0$ Lie supergroup yields the following right-invariant super one-forms:
\begin{eqnarray}\label{3.35}
R_{\pm}^1= \partial_{\pm} t + \partial_{\pm} \psi ~\frac{\psi}{2},~~~~~~R_{\pm}^2 = \partial_{\pm} \varphi,~~~~~~
R_{\pm}^3 = -\partial_{\pm} \psi,~~~~~R_{\pm}^4 = -\partial_{\pm} \chi.
\end{eqnarray}
The construction of the corresponding super Poisson structures requires the formation of the Drinfeld superdouble.
The Lie superalgebra of the semie-Abelian Drinfeld superdouble $((2{\A}_{1,1}+2{ \A})^0 , {\cal I}_{_{(2|2)}})$ is characterized by the following non-vanishing (anti-)commutation relations
\begin{eqnarray}\label{3.36}
\{T_{_3} , T_{_3}\} = T_{_1},~~~~~~~~~~~[T_{_3} , {\tilde T}^1]=-{\tilde T}^3.
\end{eqnarray}
Furthermore, we maintain the spectator-dependent matrices subject to the same condition \eqref{3.11}.
Consequently, the line element and $B$-field of the original $\sigma$-model are expressed as follows:
\begin{align}
ds^2 =& \frac{1}{1-\frac{2M}{r}} ~ dr^2   - (1-\frac{2M}{r}) ~ {d t}^2 +  r^{2} ({d \theta}^2+ \sin^2 \theta ~ {d \varphi}^2)
+ (1-\frac{2M}{r}) \psi  ~ d t d \psi-2 F_0 ~ d\psi d\chi,\label{3.37}\\
B = &-\frac{1}{2} h (r)~ d\chi \wedge d\chi.\label{3.38}
\end{align}
One can demonstrate that this background is conformally invariant up to one-loop order, with a constant dilaton
$\Phi ={\cal C}_{_0}$ and a vanishing cosmological constant.
An important feature of this solution is that its purely bosonic part is again exactly the Schwarzschild geometry.
Interestingly, the $B$-field is exactly the same functional form as the $B$-field obtained for the original models of the previous two examples.
Nevertheless, the full supergeometry is not equivalent to the two backgrounds associated with $(C^{3} + A)$ and $C^3 \otimes { A}_{1,1}$.
The crucial distinction is the presence of the Grassmann-dependent mixed term
$(1-\frac{2M}{r}) \psi  ~ d t d \psi$,
which couples the fermionic coordinate $\psi$ directly to the time direction of the Schwarzschild geometry.
Therefore, although the bosonic geometry remains unchanged, the fermionic extension of spacetime is non-trivial
and carries explicit information about the underlying Lie superalgebraic structure.
Consequently, the aforementioned background and the two preceding ones share a common bosonic gravitational sector and an identical $B$-field
whereas their inequivalence is encoded in the fermionic components of the supermetric and in their corresponding superisometry algebras.
The Killing structure of the $(2{ A}_{1,1}+2{ A})^0$ background provides a particularly transparent manifestation of this difference.
It admits five bosonic Killing vectors,
\begin{align}
K^B_{_1} =& \overrightarrow{\frac{\partial}{\partial t}},~~~~~~~~~~~~~~~~~~~~~~~~~~~~~~~~~~~~
K^B_{_2} = \sin \varphi  \overrightarrow{\frac{\partial}{\partial \theta}} + \cos \varphi \cot \theta \overrightarrow{\frac{\partial}{\partial \varphi}},
\nonumber\\
K^B_{_3} =& \cos \varphi  \overrightarrow{\frac{\partial}{\partial \theta}} - \sin \varphi \cot \theta \overrightarrow{\frac{\partial}{\partial \varphi}},
~~~~~~~~
K^B_{_4} = \overrightarrow{\frac{\partial}{\partial \varphi}},\nonumber
\end{align}
together with the additional bosonic generator
\begin{align}
K^B_{_5} = \psi \overrightarrow{\frac{\partial}{\partial \chi}}, \nonumber
\end{align}
and two fermionic Killing vectors,
\begin{eqnarray}
K^F_{_1} = \frac{\psi}{2} \overrightarrow{\frac{\partial}{\partial t}} +  \overrightarrow{\frac{\partial}{\partial \psi}},~~~~~~~~~~~~~
K^F_{_2} =  \overrightarrow{\frac{\partial}{\partial \chi}}.\nonumber
\end{eqnarray}
The first four bosonic generators reproduce the complete time-translation and rotational symmetry of the bosonic Schwarzschild spacetime.
The additional generator $K^B_{_5}$ acts in the fermionic directions and reflects the non-trivial structure of the fermionic part.
More importantly, the fermionic Killing vector $K^F_{_1}$ contains both a fermionic translation and a bosonic time-translation component. This mixed generator is directly correlated with the explicit $(\psi,dt,d\psi)$ coupling appearing in the supermetric.
Hence, the fermionic deformation in this background is associated with the time part rather than the angular one.
The non-vanishing graded commutation relations among these Killing supervectors are
\begin{eqnarray}\label{3.39}
&&[K^B_{_2}  , K^B_{_3}] =  K^B_{_4},~~~~~[K^B_{_3}  , K^B_{_4}] =  K^B_{_2},~~~~[K^B_{_4}  , K^B_{_2}] =  K^B_{_3},\nonumber\\
&&[K^F_{_1}  , K^B_{_5}] =  K^F_{_2},~~~~~\{K^F_{_1}  , K^F_{_1}\} = K^B_{_1}.
\end{eqnarray}
In particular, $K^B_{_1}$ commutes with all generators, reflecting the invariance of the background under time translations.
The three rotational generators $(K^B_{_2}, K^B_{_3}, K^B_{_4})$ reproduce the usual $so(3)$ Lie algebra,
reflecting the spherical symmetry of the Schwarzschild spacetime.
The relation $[K^F_{_1}  , K^B_{_5}] =  K^F_{_2}$ demonstrates the coupling between the bosonic transformation generated by $ K^B_{_5}$
and the fermionic part, whereas the anti-commutator $\{K^F_{_1}  , K^F_{_1}\} = K^B_{_1}$ shows that the square of
the fermionic Killing supervector closes onto the time-translation generator.
Indeed, the Lie superalgebra spanned by the generators $(K^B_{_1}, K^B_{_5};  K^F_{_1}, K^F_{_2})$ is none other than $({\C}^{3} + \A)$.

We now turn to the dual model constructed on the supermanifold ${\tilde {\M}} \approx O \times {\tilde { G}}$ with the Abelian Lie supergroup
${\tilde { G}} ={I}_{_{(2|2)}}$. The associated super Poisson structure is determined by the Lie superalgebra \eqref{3.36} together with the dual version of equations \eqref{2.11}. In this instance, the only non-vanishing component is ${\tilde \Pi}_{{{\tilde \psi} {\tilde \psi}}} = -{{\tilde x}_{_1}}$.
Finally, the corresponding dual background can be cast in the form
\begin{align}
{\tilde ds^2}=& \frac{1}{1-\frac{2M}{r}}  dr^2 + r^{2} ~ {d \theta}^2 -\frac{1}{1-\frac{2M}{r}}  {d {\tilde x}_{_{1}}}^2  + \frac{1}{r^{2} \sin^2 \theta} {d {\tilde x}_{_{2}}}^2 - \frac{2}{F_{0}} ~ d {\tilde \psi} d{\tilde \chi},\label{3.40}\\
{\tilde B}=& -\frac{\tilde{x_{1}}-h(r)}{2 F_{0}^{2}} d {\tilde \chi} \wedge d {\tilde \chi}.\label{3.41}
\end{align}
In the preceding examples, we observed that the non-Abelian dual pairs are characterized by their distinct $\tilde B$-fields.
Notably, while the resulting dual pairs share identical metrics, they are distinguished solely by their respective $\tilde B$-field configurations.
Consequently, this dual metric shares the same fundamental structure as the preceding duals including the singular points and the superisometry algebra, implying that the scalar and Kretschmann curvatures are governed by equations \eqref{3.22} and \eqref{3.23}, respectively.
Furthermore, using \eqref{3.41} and \eqref{3.11} together with \eqref{2.16} it can be shown that the only non-vanishing components of the field strength are
${\tilde H}_{_{r {\tilde \chi} {\tilde \chi}}}=-(1-\frac{2M}{r})^{-1}/{F_0}^2 r^2$ and ${\tilde H}_{_{{\tilde x}_{_{1}} {\tilde \chi} {\tilde \chi}}}={1}/{{F_0}^2}$.
The dual dilaton, obtained from \eqref{2.18}, is given by the same equation \eqref{3.21}.
Utilizing these results, one can verify that the field equations \eqref{2.13}-\eqref{2.15} are satisfied with a vanishing cosmological constant.

Finally, we conclude this section by exploring the similarities and structural differences among the backgrounds derived from
the three Lie supergroups, denoted as $(C^{3} + A)$, $C^{3} \otimes A_{1,1}$, and $(2{ A}_{1,1}+2{ A})^0$.
This comparative analysis highlights the universal features of the dual geometries.
\\\\
\smallskip
$\bullet$~{\bf Comparison and significance of the three original supergeometries}
\\
The super Poisson-Lie T-duality construction on the non-isomorphic Lie supergroups $(C^{3} + A)$, $C^{3} \otimes A_{1,1}$, and $(2{ A}_{1,1}+2{ A})^0$ leads to three distinct six-dimensional supergeometries with the same set of bosonic coordinates $(t,r,\theta,\varphi)$
and fermionic coordinates $(\psi,\chi)$. Although the three resulting backgrounds share the same radial, time, and angular bosonic structure
and possess an identical $B$-field, their fermionic parts and superisometry algebras are substantially different.
This provides a direct illustration of how the underlying Lie superalgebraic structure is reflected in the geometry and
symmetry content of the dual background.
The difference between the three solutions originates primarily from the different realization of the fermionic geometry and, consequently, from the different superisometry structures induced by the three non-isomorphic Lie superalgebras.
This is particularly useful from the perspective of super Poisson-Lie T-duality, since it demonstrates that inequivalent superalgebraic data can produce backgrounds sharing the same bosonic Schwarzschild limit and the same $B$-field part while exhibiting markedly different supersymmetric extensions.
A further important result is obtained by comparing curvature invariants.
Despite the inequivalence of the three supergeometries, all three backgrounds have vanishing scalar curvature,
and the same Kretschmann invariant.
This result is highly significant. The equality of the scalar curvature and Kretschmann invariant demonstrates that,
at the level of these curvature scalars, the three backgrounds possess the same gravitational curvature structure
as the Schwarzschild geometry.
However, the equality of these curvature invariants does not imply that the three supergeometries are identical.
Curvature scalars probe only particular invariant combinations of the geometry and therefore do not, by themselves, capture the complete structure of the fermionic part. The three backgrounds demonstrate precisely this point: they share the same bosonic limit, the same scalar curvature, the same Kretschmann invariant, and the same functional form of the $B$-field, but differ in their Grassmann-dependent metric components and in their super Killing structures.

Overall, these three examples provide a useful test of the sensitivity of super Poisson-Lie T-duality to the underlying Lie superalgebraic structure.
The fact that different superalgebraic inputs lead to backgrounds with the same Schwarzschild bosonic limit and identical curvature invariants,
but with different fermionic couplings and superisometry algebras, demonstrates that the fermionic extension contains non-trivial duality information
that is invisible in the purely bosonic geometry.
This observation motivates a more detailed analysis of the three dual backgrounds beyond their bosonic curvature properties.
In the following part of the analysis, we will investigate their mutual differences and the detailed properties of the dual backgrounds, including their supergeometric structures, Killing superalgebras, and other quantities relevant for distinguishing the resulting super Poisson-Lie T-dual solutions.
\\\\
\smallskip
$\bullet$~{\bf Comparison of the three dual supergeometries}
\\
It is worth emphasizing the close relationship between the three dual backgrounds obtained from the non-isomorphic
Lie supergroups $(C^{3} + A)$, $C^{3} \otimes A_{1,1}$, and $(2{ A}_{1,1}+2{ A})^0$. Although the three supergroups are algebraically distinct, their corresponding dual backgrounds exhibit a remarkable degree of similarity.
In particular, the three dual metrics are exactly identical. Consequently, all geometric quantities that depend solely on the metric, including the scalar curvature and the Kretschmann invariant, have the same functional form in the three backgrounds.
Therefore, the location and nature of the curvature singularities are identical in all three models.
The same conclusion applies to the superisometries: since the metric is common to the three backgrounds, their Killing supervector fields
and the associated bosonic and fermionic isometry structures are equivalent.
Thus, from the purely metric and isometric points of view, the three dual models cannot be distinguished.
Hence, the one-loop conformal properties associated with the metric and dilaton parts are shared by the three dual backgrounds.
This agreement is particularly interesting because the three backgrounds originate from non-isomorphic supergroup structures.
It indicates that the algebraic distinction between the three Drinfeld superdouble decompositions does not necessarily lead to a distinction in the purely bosonic geometric part or in the one-loop dilaton required for conformal invariance.
The essential difference between the three backgrounds appears instead in their $\tilde B$-fields.
The $\tilde B$-field enters the worldsheet action independently and contributes to the corresponding flux and fermionic interactions.
Therefore, the non-trivial difference in the fermionic $\tilde B$-field can encode physical information that is
invisible to curvature invariants and super Killing symmetries.
Therefore, the comparison of dual backgrounds resulting from super Poisson-Lie T-duality should be carried out at the level of the complete set of background fields, rather than merely through the metric part.

In the subsequent section, we apply the super Poisson-Lie T-plurality formulated in \cite{Eghbali3} (see, also, \cite{vonUnge})
to derive chains of the conformal duality/plurality for four-dimensional string backgrounds incorporating a Schwarzschild spacetime coupled to two fermionic fields.

\section{Super Poisson-Lie T-plurality in $\sigma$-models with the Schwarzschild metric coupled to two fermionic fields}
\label{Sec.IV}

In the previous section, we considered the super non-Abelian T-dualization of the Schwarzschild geometry in the presence of two fermionic fields.
The analysis there was formulated in terms of semi-Abelian superdoubles whose associated Lie superalgebras are mutually non-isomorphic.
Here, we turn to a more general framework. In particular, the ${\C}^3 \oplus {\A}_{1,1}$ and $({\C}^3 + {\A})$ Lie superalgebras
admit different Manin supertriple decompositions \cite{ER15,Eghbali3}.
These alternative decompositions can be carried out within the corresponding Drinfeld superdoubles.
This observation gives the appropriate setting to examine the super Poisson-Lie T-plurality transformations.


\subsection{A brief review of the super Poisson-Lie T-plurality}
\label{IV.1}

For completeness, we recall the basic construction of super Poisson-Lie T-plural $\sigma$-models on Lie supergroups \cite{Eghbali3}.
Consider an untransformed $\sigma$-model defined on the supergroup $G$ with element $g \in G$, which forms the first sub-supergroup of the Drinfeld superdouble $D=(G , {\tilde G})$. The corresponding action can be expressed as
\begin{eqnarray}\label{4.1}
S = \frac{1}{2} \int_{_{\Sigma}} d\sigma^{+}  d\sigma^{-} \Big[(-1)^a  R_{+}^a(g)~{{E}_{_{ab}}}(g, y^i)~
 R_{-}^b(g) + \partial_{+} y^{i} F_{_{ij}}(y^i) \partial_{-} y^{j} \Big],
\end{eqnarray}
where the coupling $F_{_{ij}}(y^i)$ may be at most function of the orbit
variables $y^{i}$ only, and the background matrix ${{E}_{_{ab}}}(g, y^i)$ is defined as in \eqref{2.10},
\begin{eqnarray}\label{4.2}
{E_{_{ab}}} = {\big(E^{{-1}}_{0}+ \Pi\big)_{{ab}}}^{\hspace{-3mm}-1}.
\end{eqnarray}

Suppose now that the same Drinfeld superdouble allows more than one Manin supertriple decomposition.
Let  $\mathbb{X}_{_A}=\{T_{_a} , {\tilde T}^{^b}\}$, with $a,b=1,\ldots,\dim G$, denote the generators of the Lie sub-superalgebras $\G$ and $\tG$ associated with the original $\sigma$-model \eqref{4.1}.
We introduce another basis, $\mathbb{X}'_{_A}=\{U_{_a} , {\tilde U}^{^b}\}$,
corresponding to a second Manin supertriple $({\G}_{_U} , {\tG}_{_U})$ embedded in the same Drinfeld superdouble, so that $\D \cong {\D}_{_U}$.
Both decompositions obey the defining relations \eqref{2.3} and \eqref{2.4}.
The two superdouble descriptions are regarded as isomorphic when a superinvertible matrix ${C_{_{A}}}^{B}$ exists such that the transformation
$\mathbb{X}_{_A} = (-1)^{^B}~ {C_{_{A}}}^{B}~ \mathbb{X}'_{_B}$
maps the generators of one superdouble into those of the other while leaving the invariant bilinear form $\big<. , .\big>$ unchanged.
In terms of the components, this change of basis takes the form \cite{Eghbali3}
\begin{eqnarray}
T_{_a} &=& (-1)^{^c}~ {F_{_a}}^{^c}~U_{_c} + G_{_{ac}}~{\tilde U}^{^c}, \nonumber\\
{\tilde T}^{^b} &=& (-1)^{^c}~ H^{^{bc}}~U_{_c} + {K^{^b}}_{_{c}}~{\tilde U}^{^c}.\label{4.3}
\end{eqnarray}
Using the generators $U_{_a}$, we may consequently construct a $\sigma$-model on the corresponding Lie supergroup, with supergroup element denoted by $g_{_U}$.
The resulting super plurality-transformed action retains the structure of \eqref{4.1}, while its background matrix is now evaluated on $g_{_U}$. Hence,
\begin{eqnarray}\label{4.4}
S_{_U} = \frac{1}{2} \int_{_{\Sigma}} d\sigma^{+}  d\sigma^{-} \Big[(-1)^a  R_{+}^a(g_{_U})~{{E}_{_{ab}}}(g_{_U}, y^i)~
 R_{-}^b(g_{_U}) + \partial_{+} y^{i} F_{_{ij}}(y^i) \partial_{-} y^{j} \Big],
\end{eqnarray}
where the transformed background is determined by
\begin{eqnarray}\label{4.5}
E(g_{_{_U}} , y^i)=\big(N M^{^{-1}} + \Pi(g_{_{_U}})\big)^{^{-1}},
\end{eqnarray}
in which the matrices $M, N$, and $\Pi(g_{_{_U}})$ in the form of their components may be expressed as
\begin{eqnarray}
M_{_{ab}}&=& (-1)^{^c}~ {{(K^{st})}_{_{a}}}^{{c}}~ {E_{_0}}_{_{cb}} - {(G^{st})}_{_{ab}},\label{4.6}\\
{N^{^a}}_{_{b}}~&=& {(F^{st})^{a}}_{_{b}}- (-1)^{^c}~ {(H^{st})}^{^{ac}}~ {E_{_0}}_{_{cb}},\label{4.7}\\
\Pi^{^{ab}}(g_{_{_U}})&=& (-1)^{^c}~ b^{^{ac}}(g_{_{_U}}) ~ {{(a^{-1})}_{_c}}^{^b} (g_{_{_U}}).\label{4.8}
\end{eqnarray}
Here, the sub-matrices $a(g_{_{_U}})$ and $b(g_{_{_U}})$ are defined in the same manner as in \eqref{2.11}.

An additional contribution must be taken into account when implementing the plurality transformation at the quantum level.
As demonstrated by von Unge in Ref. \cite{vonUnge}, the path-integral treatment generates a correction associated with integrating out the fields
belonging to the dual group $\tilde G$.
Under appropriate circumstances, this contribution can be incorporated into the one-loop transformation of the dilaton field.
Following \cite{vonUnge}, for a Lie supergroup, the corresponding dilaton transformation can be written as follows:
\begin{eqnarray}\label{4.9}
\Phi_{_{_U}} = \phi^{^{(0)}} +\frac{1}{2}\log \big|s\hspace{-0.6mm}\det\big({_{_a}}{E_{_b}}(g_{_{_U}})\big) \big|-
\frac{1}{2}\log \big|s\hspace{-0.6mm}\det\big({_{_a}}{M_{_b}}\big) \big|+
\frac{1}{2}\log \Big|s\hspace{-0.6mm}\det\Big({_{_a}}{(a^{-1})^{^b}} {(g_{_{_U}})}\Big)\Big|,
\end{eqnarray}
where $\phi^{^{(0)}}$ denotes the dilaton that renders the initial $\sigma$-model conformal.
In general, this quantity may depend on the coordinates parametrizing $G_{_U}$.
The usual super Poisson-Lie T-duality is recovered as a particular limit of the above plurality transformation. Indeed, taking
$F=K={\bf 1}$ and $G=H=0$ in \eqref{4.3} gives ${N^{^a}}_{_{b}} = {\delta^{^a}}_{_{b}}$ and $M_{_{ab}} = {E_{_{0}}}_{ab}$.
Substitution of these expressions into \eqref{4.9} reproduces the dilaton transformation in \eqref{2.17}.

We will use this formulation in the rest of this section to study the super Poisson-Lie T-plurality transformations of the Schwarzschild $\sigma$-model coupled to two fermionic fields
on some of the Manin supertriples, inside the related Drinfeld superdoubles.


\subsection{Conformal supersymmetric deformations via super Poisson-Lie T-plurality with respect to the ${C}^3 \otimes {A}_{1,1}$ Lie supergroup}
\label{IV.2}

Recently in Ref. \cite{Meysam} the general Lie superbialgebra structures on the ${\C}^3 \oplus {\A}_{1,1}$ Lie superalgebra were obtained by solving the super Jacobi identities.
As a result, many real $(4|4)$-dimensional Drinfeld superdoubles were obtained which decompose into a class of isomorphic Manin supertriples listed below
\begin{eqnarray}\label{4.10}
&&~~\big({\C}^3 \oplus {\A}_{1,1} , {\cal I}_{_{(2|2)}}\big) \cong  \big({\C}^3 \oplus {\A}_{1,1} , {\C}^3 \oplus {\A}_{1,1}.i\big)
\cong \big({\C}^3 \oplus {\A}_{1,1} , {(2{\A}_{1,1}+2{ \A})^0}.i\big) \nonumber\\
&&~~~~~~~~~~~~~~~~~~~~~~~~~\cong  \big({\C}^3 \oplus {\A}_{1,1} , (({\A}_{1,1}+2{ \A})^2 \oplus {\A}_{1,1}).i\big).
\end{eqnarray}
The isomorphic transformations between these superdoubles have been presented in \cite{Meysam}.
In the following, we use these transformations, together with the super Poisson-Lie T-plurality formulation mentioned above, to construct the models.

\subsubsection*{{\it \underline{Super Poisson-Lie T-plurality with respect to
the superdouble $\D={\bf \big({\C}^3 \oplus {\A}_{1,1} , {\cal I}_{_{(2|2)}}\big)}$}}}
\smallskip
In order to obtain a possible conformal duality chain on the isomorphic Drinfeld superdoubles of \eqref{4.10}
we begin with $\D=({\C}^3 \oplus {\A}_{1,1} , {\cal I}_{_{(2|2)}})$
whose Lie superalgebra has been given in \eqref{3.26}.
In Sec. \ref{Sec.III}, we derived the super one-forms $ R_{\pm}^a(g)$, as shown in \eqref{3.27}, and $\Pi(g)$ on the ${C}^3 \otimes {A}_{1,1}$, parameterized by the supercoordinates $(t, \varphi; \psi, \chi)$.
It was found that $\Pi(g) = 0$, which gives us $E = E_0$  as expected for the canonical decomposition.
Considering the coordinates $y^i =(r, \theta)$ we choose the coupling matrices in the following form
{\small \begin{eqnarray}
{E_{0}}_{_{ab}}=\left( \begin{tabular}{cccc}
                 $-(1-\frac{2M}{r})$ & 0 & 0 & 0 \\
                 0 & $r^{2} \sin^2 \theta$  & 0& 0 \\
                 0 & 0& 0 & $F_0$ \\
                 0 & 0 & -$F_0$ & $h(r)$ \\
                 \end{tabular} \right),~ ~~~~~~~~~~~
                 F_{i j} = \left( \begin{tabular}{cccc}
                 $(1-\frac{2M}{r})^{-1}$ & 0 \\
                 0 & $r^{2}$ \\
                 \end{tabular} \right),\label{4.11}
\end{eqnarray}}
with $h(r)$ defined by \eqref{3.11}. Thus, the action \eqref{4.1} can be expressed on the superdouble $\D=({\C}^3 \oplus {\A}_{1,1} , {\cal I}_{_{(2|2)}})$.
The corresponding line element and $B$-field are then read
\begin{align}
ds^2 =& \frac{1}{1-\frac{2M}{r}} ~ dr^2 - (1-\frac{2M}{r}) ~ {d t}^2 + r^{2} ({d \theta}^2 +\sin^2 \theta ~ {d \varphi}^2)
-2 F_0 ~ d\psi d\chi,\label{4.12} \\
B =& -\frac{1}{2} h(r)~ d\chi \wedge d\chi,\label{4.13}
\end{align}
which is nothing but the background given by \eqref{3.28} and \eqref{3.29}.
Additionally, in Sec. \ref{Sec.III}, we demonstrated that the dilaton field $\Phi = \phi^{^{(0)}} = {\cal C}_{_0}$
ensures the one-loop conformal invariance of the background.


\subsubsection*{{\it \underline{Super Poisson-Lie T-plurality with respect to
the superdouble ${\D}_{_U}={\bf \big({\C}^3 \oplus {\A}_{1,1} , {\C}^3 \oplus {\A}_{1,1}.i\big)}$}}}
\smallskip
The Lie superalgebra of the superdouble ${\D}_{_U}=\big({\C}^3 \oplus {\A}_{1,1} , {\C}^3 \oplus {\A}_{1,1}.i\big)$
satisfies the following set of non-trivial (anti-)commutation relations:
\begin{align}
[U_{_1} , U_{_4}] =&U_{_3},~~~~~~~~~[{\tilde U}^{^2} , {\tilde U}^{^3}]= {\tilde U}^{^4},~~~~~~~~~~~{[U_{_1} , {\tilde U}^{^3}]}= -{\tilde U}^{^4},\nonumber\\
{[U_{_4} , {\tilde U}^{^2}]}=&  U_{_3},~~~~~~~~\{U_{_4} , {\tilde U}^{^3}\}= U_{_2}-{\tilde U}^{^1},\label{4.14}
\end{align}
where $U_{_a}$'s denote the generators of the sub-superalgebra ${\C}^3 \oplus {\A}_{1,1}$,
while ${\tilde U}^a$'s represent the generators of the dual sub-superalgebra ${\C}^3 \oplus {\A}_{1,1}.i$.
Here and henceforth we denote the bosonic bases by $(U_{_1}, U_{_2}, {\tilde U}^{^1}, {\tilde U}^{^2})$, and fermionic ones by
$(U_{_3}, U_{_4}, {\tilde U}^{^3}, {\tilde U}^{^4})$.
The isomorphism between the superdoubles $\D=({\C}^3 \oplus {\A}_{1,1} , {\cal I}_{_{(2|2)}})$
and ${\D}_{_U}=({\C}^3 \oplus {\A}_{1,1} , {\C}^3 \oplus {\A}_{1,1}.i)$ is given by the following transformation:
\begin{align}
{T_{_1}} =&U_{_1},~~~~~~~~~~~~~~~~{\tilde T}^{^1} =- U_{_2} + {\tilde U}^{^1},\nonumber\\
{T_{_2}} =&U_{_2},~~~~~~~~~~~~~~~~{\tilde T}^{^2} = U_{_1} + {\tilde U}^{^2},\nonumber\\
{T_{_3}} =& - U_{_3},~~~~~~~~~~~~{\tilde T}^{^3} =- {\tilde U}^{^3},\nonumber\\
{T_{_4}} =& - U_{_4},~~~~~~~~~~~~{\tilde T}^{^4} =- {\tilde U}^{^4}.\label{4.15}
\end{align}
A comparison of equations \eqref{4.3} and \eqref{4.15} yields the sub-matrices ${F_{_a}}^{^c},  G_{_{ac}},  H^{^{bc}}$ and ${K^{^b}}_{_{c}}$.
Substituting these, along with $E_{0}$ (from \eqref{4.11}) and the relations \eqref{4.6} and \eqref{4.7}, allows us to
calculate the matrices $M_{_{ab}}$ and ${N^{^a}}_{_{b}}$, resulting in
\begin{eqnarray}
NM^{^{-1}} = \left( \begin{array}{cccc}
\frac{-1}{1-{\frac{2M}{r}}}  & -1  & 0 & 0\\
1  & \frac{1}{r^{2} \sin^2 \theta}  & 0 & 0\\
0 & 0  & \frac{-h(r)}{{F_{0}}^2} & \frac{1}{F_{0}}\\
0  & 0  & \frac{-1}{F_{0}} & 0\\
\end{array} \right).\label{4.16}
\end{eqnarray}
Now, we use the parametrization of a group element $g_{_U} \in {C}^3 \otimes {A}_{1,1}$ as in \eqref{3.5}, but with
$T_{_a}$'s replaced by $U_{_a}$'s. Then, one utilizes
formula \eqref{2.11} to calculate the matrices $a(g_{_U})$ and $b(g_{_U})$ for this decomposition, giving us
\begin{eqnarray}\label{4.17}
a_{_{a}}^{^{^{~b}}}(g_{_U}) =\left( \begin{array}{cccc}
1  & 0  & -\chi & 0\\
0  & 1  & 0 & 0\\
0 & 0  & -1 & 0\\
0  & 0  & {x_{_{1}}} & -1\\
\end{array} \right),~~~~~~~~~~b^{^{ab}}(g_{_U})=\left( \begin{array}{cccc}
0  & 0  & 0 & 0\\
0  & 0 &  \chi & 0\\
0 & -\chi  & 0 & 0\\
0  & 0  & 0 & 0\\
\end{array} \right).
\end{eqnarray}
Substituting \eqref{4.17} into \eqref{4.8}, one finds that the only non-vanishing component of the super Poisson structure is given by
${\Pi}^{{{x_{_{2}}} \psi}}(g_{_U}) = \chi$. Consequently, these results determine the coupling matrix
\begin{eqnarray}
{E}_{_{ab}}(g_{_U}) = \left( \begin{array}{cccc}
\frac{1-{\frac{2M}{r}}}{\Delta} & \frac{\Delta+1}{\Delta}  & 0 & \frac{F_{0}(\Delta+1)}{\Delta} \chi \\
-\frac{\Delta+1}{\Delta}  & -\frac{r^{2} \sin^2 \theta}{\Delta}  & 0 & -\frac{F_{0} r^{2} \sin^2 \theta}{\Delta} \chi\\
0 & 0  & 0 & F_{0}\\
\frac{F_{0}(\Delta+1)}{\Delta} \chi  & \frac{F_{0} r^{2} \sin^2 \theta}{\Delta} \chi  & -F_{0} & h(r)\\
\end{array} \right),\label{4.18}
\end{eqnarray}
where $\Delta: = r^{2} \sin^2 \theta~ (1-{\frac{2M}{r}})-1 $.
Using \eqref{4.18} and the super one-forms on the ${C}^3 \otimes {A}_{1,1}$ of equation
\eqref{3.27}, one can write the action of model on the superdouble $({\C}^3 \oplus {\A}_{1,1} , {\C}^3 \oplus {\A}_{1,1}.i)$
in the form of \eqref{4.4}.
Finally,  background including the line element and $B$-field is, in the coordinate basis, read off
\begin{align}
ds^2 =& \frac{1}{1-{\frac{2M}{r}}} dr^2 + r^2 {d{\theta}}^2 +\frac{1-{\frac{2M}{r}}}{\Delta} d{x_{1}}^2  -\frac{r^{2} \sin^2 \theta}{\Delta} d{x_{2}}^2
- \frac{2 F_{0}(\Delta+1)}{\Delta} \chi~ d{x_{1}}  d \chi -2 F_{0} ~d \psi  d \chi,  \nonumber\\
B=&\frac{\Delta+1}{\Delta}d{x_{1}} \wedge d{x_{2}} +  \frac{F_{0} r^{2} \sin^2 \theta}{\Delta}\chi ~ d{x_{2}} \wedge  d \chi - \frac{1}{2} h(r)~ d \chi \wedge  d \chi.\label{4.19}
\end{align}
Compared to the backgrounds presented in the previous section, the new background differs more fundamentally in
structure regarding both the bosonic part and the boson-fermion interactions.
In the background \eqref{4.19}, however, the bosonic part incorporates non-trivial factors dependent on $\Delta$,
so that the coefficients of $dx_1^2$ and $dx_2^2$ no longer appear in the standard Schwarzschild form.
Furthermore, the $B$-field exhibits a richer structure; in addition to the fermionic term $-\frac{1}{2} h(r)\,d\chi\wedge d\chi$, it includes a purely bosonic term $\frac{\Delta+1}{\Delta}dx_1\wedge dx_2$ as well as a fermion-boson term proportional to $\chi\,dx_2\wedge d\chi$.
Thus, unlike the previous backgrounds, the effect of the duality is manifested not only through the fermionic directions but also
directly in the bosonic structure and $B$-field.
Another distinctive feature of this new background is the boson-fermion coupling term
$-\frac{2F_0(\Delta+1)}{\Delta}\,\chi\,dx_1\,d\chi $ which differs significantly from the previous cases.
In this new supergeometry, the coupling with \(x_1\) is mediated by \(\chi\), and its coefficient depends non-trivially on \(r\) and \(\theta\) via \(\Delta\); thus, this term not only links the fermionic structure to the bosonic part but also renders this linkage dependent on the underlying geometry.
Consequently, this background can be viewed as a richer supergeometric structure wherein the distinctions arising from super Poisson-Lie duality are not confined solely to the fermionic part or supersymmetries but manifest simultaneously in the bosonic components of the metric, the \(B\)-field, and the coupling between bosonic and fermionic coordinates.
Determining the Killing supervectors of this supergeometry and comparing its superisometry algebra with that of the previous cases could constitute another crucial aspect of the distinction among these backgrounds.

Here and henceforth we omit the analysis of the scalar curvature, the Kretschmann invariant, and the superisometry structure associated with the metrics.
Our analysis is confined to verifying the one-loop conformal invariance of the backgrounds, incorporating the dilaton solution from \eqref{4.9}.
In order to calculate the corresponding dilaton field we find that
$s\hspace{-0.6mm}\det(_a{{E}_{_{b}}(g_{_U})}) = \frac{\Delta+1}{{F_{0}}^2 \Delta},~s\hspace{-0.6mm}\det(_a{{M}_{_{b}}(g_{_U})}) =-\frac{\Delta+1}{{F_{0}}^2}$, and
$s\hspace{-0.6mm}\det(a_{_{a}}^{{{~b}}}(g_{_U})) =1$. Then, the dilaton is obtained by making use of
\eqref{4.9}, and by remembering that $\phi^{^{(0)}}= {\cal C}_{_{0}}$ one gets the final result in the form
\begin{eqnarray}\label{4.20}
\Phi_{_U} = {\cal C}_{_{0}} -\frac{1}{2} \log\big|\Delta\big|.
\end{eqnarray}


\subsubsection*{{\it \underline{Super Poisson-Lie T-plurality with respect to
the superdouble ${\D}_{_U}={\bf \big({\C}^3 \oplus {\A}_{1,1} , {(2{\A}_{1,1}+2{ \A})^0}.i\big)}$}}}
\smallskip
From the Manin supertriples of \eqref{4.10} we see that $\D=({\C}^3 \oplus {\A}_{1,1} , {\cal I}_{_{(2|2)}})$ and
${\D}_{_U}={\big({\C}^3 \oplus {\A}_{1,1} , {(2{\A}_{1,1}}}$\\${{+2{ \A})^0}.i\big)}$ as Lie
superalgebras are isomorphic, and so one can find an isomorphism transforming the Lie multiplication of $\D$ into that of ${\D}_{_{U}}$
and preserving the canonical form of the bilinear form $\big<. , .\big>$ such that they belong to the same Drinfeld superdouble.
The isomorphism transformation of Manin supertriples between $({\C}^3 \oplus {\A}_{1,1} , {\cal I}_{_{(2|2)}})$ and
${\big({\C}^3 \oplus {\A}_{1,1} , {(2{\A}_{1,1}+2{ \A})^0}.i\big)}$ is given by
\begin{align}
{T_{_1}} =&U_{_1},~~~~~~~~~~~~~~~~{\tilde T}^{^1} = {\tilde U}^{^1},\nonumber\\
{T_{_2}} =&U_{_2},~~~~~~~~~~~~~~~~{\tilde T}^{^2} ={\tilde U}^{^2},\nonumber\\
{T_{_3}} =& - U_{_3},~~~~~~~~~~~~{\tilde T}^{^3} =-\frac{\eta}{2} U_{_4}- {\tilde U}^{^3},\nonumber\\
{T_{_4}} =& - U_{_4},~~~~~~~~~~~~{\tilde T}^{^4} =-\frac{\eta}{2} U_{_3} - {\tilde U}^{^4},\label{4.21}
\end{align}
where $\{U_{_a} , {\tilde U}^{^b}\}$ are elements of bases of the superdouble
${\big({\C}^3 \oplus {\A}_{1,1} , {(2{\A}_{1,1}+2{ \A})^0}.i\big)}$ whose Lie superalgebra  is defined by
the following non-vanishing (anti-)commutation relations:
\begin{eqnarray}
[U_{_1} , U_{_4}] = U_{_3},~~~~~~\{{\tilde U}^{^3} , {\tilde U}^{^3}\}=\eta {\tilde U}^{^1},~~~~{[U_{_1} , {\tilde U}^{^3}]}=-\eta U_{_3} -{\tilde U}^{^4},~~~~~\{U_{_4} , {\tilde U}^{^3}\}=-{\tilde U}^{^1},\label{4.22}
\end{eqnarray}
where $\eta$ is a non-zero parameter.
The matrix $a_{_{a}}^{{{~b}}}(g_{_U})$ follows the same structure as presented in \eqref{4.17}.
Upon calculating $M_{_{ab}}$, ${N^{^a}}_{_{b}}$, and ${\Pi}^{^{ab}}(g_{_U})$ for the decomposition \eqref{4.21}, one obtains
\begin{eqnarray}
NM^{^{-1}} = \left( \begin{array}{cccc}
\frac{-1}{1-{\frac{2M}{r}}}  & 0  & 0 & 0\\
1  & \frac{1}{{r^{2} \sin^2 \theta}}  & 0 & 0\\
0 & 0  & \frac{-h(r)}{{F_{0}}^2} & \frac{1}{F_{0}}-\frac{\eta}{2}\\
0  & 0  & -\frac{1}{F_{0}}-\frac{\eta}{2} & 0\\
\end{array} \right),~~{\Pi}(g_{_U}) = \left( \begin{array}{cccc}
0  & 0  & 0 & 0\\
0  &0  & 0 & 0\\
0 & 0  & -\eta t & 0\\
0  & 0  & 0& 0\\
\end{array} \right),~~~\label{4.23}
\end{eqnarray}
yielding a background of the form
\begin{eqnarray}\label{4.24}
{E}_{_{ab}}(g_{_U}) &=& \left( \begin{array}{cccc}
-(1-\frac{2M}{r}) & 0      & 0 & 0\\
0     & r^{2} \sin^2 \theta & 0 & 0\\
0     & 0      & 0 & \frac{2F_{0}}{\eta F_{0} + 2}\\
0     & 0      & \frac{2F_{0}}{\eta F_{0} - 2} & -\frac{4(\eta {F_{0}}^2 t + h(r))}{{\eta}^2 {F_{0}}^2 - 4}\\
\end{array} \right).
\end{eqnarray}
Since the first sub-superalgebra of ${\D}_{_U}$ is ${\C}^3 \oplus {\A}_{1,1}$,
the right-invariant super one-forms have the same forms as in \eqref{3.27}.
Thus, using \eqref{4.4} one can construct the action of the model on the superdouble
$\big({\C}^3 \oplus {\A}_{1,1} , {(2{\A}_{1,1}+2{ \A})^0}.i\big)$
whose background is
\begin{align}
ds^2 =& \frac{1}{1-{\frac{2M}{r}}} dr^2  - (1-{\frac{2M}{r}}) ~ d{t}^2  + r^{2} ( d{\theta}^2 + \sin^2 \theta~ d{\varphi}^2 )
- \frac{8 F_{0}}{{\eta}^2 F_0^2 -4} ~d \psi  d \chi ,  \nonumber\\
B =&\frac{2 \eta {F_{0}}^2}{{\eta}^2 {F_{0}}^2 -4} d \psi \wedge d \chi -  \frac{2 h (r)}{{\eta}^2 {F_{0}}^2 -4} ~ d \chi \wedge  d \chi.\label{4.25}
\end{align}
By performing a rescaling of the coefficient $\frac{4}{{\eta}^2 F_0^2 -4}$ to unity in the present background,
we precisely recover the ${C}^3 \otimes {A}_{1,1}$ non-Abelian T-dual background, as described by equations
\eqref{3.28} and \eqref{3.29}.
It is worth noting that because the coefficient multiplying $d \psi \wedge d \chi$ is constant,
the total derivative term appearing in the $B$-field is trivial and may be ignored.
To evaluate the total dilaton contribution we use $\phi^{^{(0)}}= {\cal C}_{_{0}}$
which gives the final result, $\Phi_{_U} = {\cal C}_{_{0}}+\frac{1}{2} \log|1-\frac{{\eta}^2 {F_{0}}^2}{4}|$.
Finally, looking at one-loop beta-function equations
one can verify the conformal invariance conditions of the background \eqref{4.25} with the constant dilaton field $\Phi_{_U}$ and a
vanishing cosmological constant.


\subsubsection*{{\it \underline{Super Poisson-Lie T-plurality with respect to
the ${\D}_{_U}={\bf \big({\C}^3 \oplus {\A}_{1,1} , \big(({\A}_{1,1}+2{\A})^2 \oplus {\A}_{1,1}\big).i\big)}$}}}
\smallskip
As the last example of this subsection we ask the question of super Poisson-Lie T-plurality of the Schwarzschild spacetime coupled to two fermionic fields
with respect to the superdouble $\big({\C}^3 \oplus {\A}_{1,1} , \big(({\A}_{1,1}+2{\A})^2 \oplus {\A}_{1,1}\big).i\big)$ whose Lie superalgebra
is defined by the following non-vanishing Lie super brackets:
\begin{align}
[U_{_1} , U_{_4}] =&U_{_3},~~~~~~~~~~~\{{\tilde U}^{^3} , {\tilde U}^{^4}\}=q {\tilde U}^{^1},~~~~~~~~~{[U_{_1} , {\tilde U}^{^3}]}= - q U_{_4}
-{\tilde U}^{^4},\nonumber\\
{[U_{_1} , {\tilde U}^{^4}]}=& -q U_{_3},~~~~~~~\{U_{_4} , {\tilde U}^{^3}\}=-{\tilde U}^{^1},\label{4.26}
\end{align}
where $q$ is a non-zero parameter and, as we shall see, appears as a deformation parameter in the background of the model.
The isomorphism that transform the Lie superalgebra $\D=({\C}^3 \oplus {\A}_{1,1} , {\cal I}_{_{(2|2)}})$ into that of
${\D}_{_{U}} = \big({\C}^3 \oplus {\A}_{1,1} , \big(({\A}_{1,1}+2{\A})^2 \oplus {\A}_{1,1}\big).i\big)$
and preserving the canonical form of the bilinear form $\big<. , .\big>$,
is given by
\begin{align}
{T_{_1}} =&U_{_1},~~~~~~~~~~~~~~~~{\tilde T}^{^1} = {\tilde U}^{^1},\nonumber\\
{T_{_2}} =&U_{_2},~~~~~~~~~~~~~~~~{\tilde T}^{^2} ={\tilde U}^{^2},\nonumber\\
{T_{_3}} =& - U_{_3},~~~~~~~~~~~~{\tilde T}^{^3} =- {\tilde U}^{^3},\nonumber\\
{T_{_4}} =& - U_{_4},~~~~~~~~~~~~{\tilde T}^{^4} =-{q} U_{_4} - {\tilde U}^{^4}.\label{4.27}
\end{align}
After calculating the sub-matrices ${F_{_a}}^{^c},  G_{_{ac}},  H^{^{bc}}$ and  ${K^{^b}}_{_{c}}$, we employ formulae \eqref{4.6} and \eqref{4.7}
to obtain the matrices $M_{_{ab}}$ and ${N^{^a}}_{_{b}}$, in such a way that one must also use the matrix $E_{0}$ of equation \eqref{4.11}.
In addition, the ${\Pi}(g_{_U})$ is obtained for this decomposition by using \eqref{4.8}.
It then follows from \eqref{4.5} that
\begin{eqnarray}\label{4.28}
{E}_{_{ab}}(g_{_U}) &=& \left( \begin{array}{cccc}
 -(1-\frac{2M}{r}) & 0  & 0 &0\\
0 & r^2 {\sin^2 \theta}  & 0 & 0\\
0  & 0  & -\frac{{F_{0}}^2}{F_{0} t-h (r)} & \frac{{F_{0}}^2 t}{F_{0} t- h(r)} \\
0  & 0  & \frac{{F_{0}}(1+ q F_{0} t)}{q (F_{0} t-h(r))} & - \frac{q {F_{0}}^2 {t}^2+h(r)}{q (F_{0} t-h(r))}\\
\end{array} \right).
\end{eqnarray}
Finally, background including the line element and $B$-field takes the following form
\begin{align}
ds^2=& \frac{1}{1-{\frac{2M}{r}}} dr^2  - (1-{\frac{2M}{r}}) ~ d{t}^2  + r^{2} ( d{\theta}^2 + {\sin^2 \theta} ~ d{\varphi}^2 )
+ \frac{F_{0}}{{q}(F_{0} t - h(r) )} ~ d \psi  d \chi ,\nonumber\\
B=& \frac{{F_{0}}^2}{2(F_{0} t - h(r))} d \psi \wedge d \psi - \frac{{F_{0}}}{2 q (F_{0} t - h(r))} d \psi \wedge d \chi -  \frac{1}{2{q}} ~ d \chi \wedge  d \chi.\label{4.29}
\end{align}
The coefficient of the $d \psi  d \chi$ term in the metric depends explicitly on both the time coordinate $t$ and the radial coordinate through $h(r)$.
Thus, although the Schwarzschild bosonic geometry is retained, the fermionic part is now coupled to a dynamical scalar factor
depending on the spacetime coordinates rather than through a Grassmann-dependent deformation of a bosonic direction.
We believe that the coordinate-dependent factor, namely $\frac{1}{{q}(F_{0} t - h(r) )}$, could alter the admissible super Killing structures.
From this perspective, the above background represents a distinct class of Poisson-Lie T-dual supergeometries in which the Schwarzschild
spacetime is preserved, while the fermionic part and $B$-field acquire an explicit spacetime-dependent structure.
However, a direct calculation of its Killing supervectors and the resulting graded commutation relations enables
a meaningful algebraic comparison with previously obtained backgrounds.

Furthermore, one can verify that the model is conformally invariant at one-loop order, with a vanishing cosmological constant and a dilaton field
${\Phi}_{_U} = {\cal C}_{_0}+\frac{1}{2} \log\big| q (F_{0} t - h(r))\big|$, as derived from \eqref{4.9}.

\subsection{Conformal supersymmetric deformations via super Poisson-Lie T-plurality with respect to the $({C}^3 + {A})$ Lie supergroup}
\label{IV.3}

The classification of real $(4|4)$-dimensional Drinfeld superdoubles generated by the $({\C}^3 + {\A})$ Lie superbialgebras \cite{ER14},
along with their decomposition into 24 non-isomorphic Manin supertriples, is detailed in Ref. \cite{ER15}.
In this work, we focus exclusively on the first class, which consists of the following three isomorphic Manin supertriples:
\begin{eqnarray}\label{4.30}
\big(({\C}^3 + {\A}) , {\cal I}_{_{(2|2)}}\big) \cong  \big(({\C}^3 + {\A}) , {\C}^3 \oplus {\A}_{1,1}.i\big)
\cong \big(({\C}^3 + {\A}) , ({\C}^3 + {\A})^{\epsilon=-1}_{k=4}\big).
\end{eqnarray}

\subsubsection*{{\it \underline{Super Poisson-Lie T-plurality with respect to the $\D={\bf \big(({\C}^3 + {\A}) , {\cal I}_{_{(2|2)}}\big)}$}}}
\smallskip

In order to obtain a possible conformal duality chain on the isomorphic Drinfeld superdoubles of \eqref{4.30}
we begin with $\D=(({\C}^3 + {\A}) , {\cal I}_{_{(2|2)}})$
whose Lie superalgebra has been given in \eqref{3.4}.
To write the action \eqref{4.1} on the $\D$ we choose the coupling matrices in the same form as in \eqref{4.11}.
In addition, we need the corresponding right-invariant super one-forms, as defined in \eqref{3.6}.
The resulting background reproduces the Schwarzschild metric coupled to two fermionic fields, equation \eqref{3.9},
and $B$-field \eqref{3.10}, all subject to the conditions specified in \eqref{3.11}.


\subsubsection*{{\it \underline{Super Poisson-Lie T-plurality with respect to the ${\D}_{_U}={\bf \big(({\C}^3 + {\A}) , {\C}^3 \oplus {\A}_{1,1}.i\big)}$}}}
\smallskip

The Lie superalgebra of the superdouble $\big(({\C}^3 + {\A}) , {\C}^3 \oplus {\A}_{1,1}.i\big)$
obeys the following set of non-trivial (anti-)commutation relations:
\begin{align}
[U_{_1} , U_{_4}] =& U_{_3},~~~~~~~~~~~~\{U_{_4} , U_{_4}\} =U_{_2},~~~~~~~~~~~~~~~[{\tilde U}^{^2} , {\tilde U}^{^3}]= {\tilde U}^{^4},\nonumber\\
{[U_{_1} , {\tilde U}^{^3}]}=&-{\tilde U}^{^4},~~~~~~~~~ {[U_{_4} , {\tilde U}^{^2}]}=  U_{_3}-{\tilde U}^{^4},~~~~~~~\{U_{_4} , {\tilde U}^{^3}\}= U_{_2}-{\tilde U}^{^1},\label{4.31}
\end{align}
where $\{U_{_a}\}$ and $\{{\tilde U}^a\}$ represent the bases of the $({\C}^3 + {\A})$ and ${\C}^3 \oplus {\A}_{1,1}.i$ Lie superalgebras, respectively.
The isomorphism between the superdoubles $\big(({\C}^3 + {\A}) , {\cal I}_{_{(2|2)}}\big)$
and $\big(({\C}^3 + {\A}) , {\C}^3 \oplus {\A}_{1,1}.i\big)$ is given by the following transformation \cite{Eghbali3}:
\begin{align}
{T_{_1}} =&U_{_1},~~~~~~~~~~~~~~~~~~{\tilde T}^{^1} =- U_{_2} + {\tilde U}^{^1},\nonumber\\
{T_{_2}} =&U_{_2},~~~~~~~~~~~~~~~~~~{\tilde T}^{^2} = U_{_1} + {\tilde U}^{^2},\nonumber\\
{T_{_3}} =& - U_{_3},~~~~~~~~~~~~~~{\tilde T}^{^3} = {\tilde U}^{^3},\nonumber\\
{T_{_4}} =&  -U_{_4},~~~~~~~~~~~~~~{\tilde T}^{^4} = {\tilde U}^{^4}.\label{4.32}
\end{align}
Comparing relations \eqref{4.3} and \eqref{4.32} one can find the sub-matrices ${F_{_a}}^{^c},  G_{_{ac}},  H^{^{bc}}$ and  ${K^{^b}}_{_{c}}$.
Then, using these and the matrix $E_{0}$ of \eqref{4.11} together with formulae \eqref{4.6} and \eqref{4.7} we can find the matrices
$M_{_{ab}}$ and ${N^{^a}}_{_{b}}$ leading to
\begin{eqnarray}
NM^{^{-1}} = \left( \begin{array}{cccc}
\frac{-1}{1-{\frac{2M}{r}}}  & -1  & 0 & 0\\
1  & \frac{1}{r^{2} \sin^2 \theta}  & 0 & 0\\
0 & 0  & \frac{-h(r)}{{F_{0}}^2} & \frac{1}{F_{0}}\\
0  & 0  & \frac{-1}{F_{0}} & 0\\
\end{array} \right).\label{4.33}
\end{eqnarray}
We adapt the parametrization of a general group element of $({C}^3 + {A})$ from \eqref{3.5} by replacing $T_{_a}$'s with $U_{_a}$'s.
Subsequently, we employ formula \eqref{2.11} to calculate the matrices $a(g_{_U})$ and $b(g_{_U})$ for this decomposition.
With the resulting matrices $a(g_{_U})$ and $b(g_{_U})$, we find that the only non-vanishing component of
the super Poisson structure is ${\Pi}^{{{x_{_{2}}} \psi}}(g_{_U}) = \chi$.
These findings determine the coupling matrix ${E}_{_{ab}}(g_{_U})$ as follows:
\begin{eqnarray}
{E}_{_{ab}}(g_{_U}) = \left( \begin{array}{cccc}
\frac{1-{\frac{2M}{r}}}{\Delta} & \frac{\Delta+1}{\Delta}  & 0 & \frac{F_{0}(\Delta+1)}{\Delta} \chi \\
-\frac{\Delta+1}{\Delta}  & -\frac{r^{2} \sin^2 \theta}{\Delta}  & 0 & -\frac{F_{0} r^{2} \sin^2 \theta}{\Delta} \chi\\
0 & 0  & 0 & F_{0}\\
\frac{F_{0}(\Delta+1)}{\Delta} \chi  & \frac{F_{0} r^{2} \sin^2 \theta}{\Delta} \chi  & -F_{0} & h(r)\\
\end{array} \right),\label{4.34}
\end{eqnarray}
where ${\Delta}$ denotes the same functional form as in \eqref{4.18}. Using \eqref{4.9} and the determinants
$s\hspace{-0.6mm}\det(_a{{E}_{_{b}}(g_{_U})}) = \frac{\Delta+1}{{F_{0}}^2 \Delta},~s\hspace{-0.6mm}\det(_a{{M}_{_{b}}(g_{_U})}) =-\frac{\Delta+1}{{F_{0}}^2}$ and
$s\hspace{-0.6mm}\det(a_{_{a}}^{{{~b}}}(g_{_U})) =1$,
along with the initial condition $\phi^{^{(0)}}= {\cal C}_{_{0}}$, we arrive at the final form of the dilaton field
\begin{eqnarray}\label{4.35}
\Phi_{_U} = {\cal C}_{_{0}} -\frac{1}{2} \log\big|\Delta\big|.
\end{eqnarray}
The supersymmetric part of the matrix ${E}_{_{ab}}(g_{_U})$ gives the metric in the coordinate basis,
whereas the super anti-symmetric part of ${E}_{_{ab}}(g_{_U})$ gives the $B$-field. Thus,
the background in the coordinate basis is read off
\begin{align}
ds^2 =& \frac{1}{1-{\frac{2M}{r}}} dr^2 + r^{2} d{\theta}^2 +\frac{1-{\frac{2M}{r}}}{\Delta} d{x_{_{1}}}^2  -\frac{r^{2} \sin^2 \theta}{\Delta} d{x_{_{2}}}^2
- \frac{2 F_{0}(\Delta+1)}{\Delta} \chi~ d{x_{_{1}}}  d \chi \nonumber \\
& ~~~~~~~~~~~~~~~~~~~~~~~~~~~~~~~~~~~~~~~~~~~~~~~~~~~~~~~~~~~~~~~~~~~~~  + \frac{r^{2} \sin^2 \theta}{\Delta} \chi~ d{x_{_{2}}}  d \chi -2 F_{0} ~d \psi  d \chi,  \nonumber\\
B=&\frac{\Delta+1}{\Delta}d{x_{_{1}}} \wedge d{x_{_{2}}} - \frac{\Delta+1}{2\Delta}\chi ~ d{x_{_{1}}} \wedge  d \chi +  \frac{F_{0} r^{2} \sin^2 \theta}{\Delta}\chi ~ d{x_{_{2}}} \wedge  d \chi \nonumber\\
&~~~~~~~~~~~~~~~~~~~~~~~~~~~~~~~~~~~~~~~~~~~~~~~~~~~~~~~~~~~~~~~~~~~~~~~~~~ - \frac{1}{2} h(r)~ d \chi \wedge  d \chi.\label{4.36}
\end{align}
Looking at the one-loop beta-function equations one verifies the conformal invariance conditions of the background \eqref{4.36} with the dilaton field \eqref{4.35} and a
vanishing cosmological constant.

\subsubsection*{{\it \underline{Super Poisson-Lie T-plurality with respect to the ${\D}_{_U}={\bf \big(({\C}^3 + {\A}) , ({\C}^3 + {\A})^{\epsilon = -1}_{k=4}\big)}$}}}
\smallskip

The Lie superalgebra of the superdouble ${ \big(({\C}^3 + {\A}) , ({\C}^3 + {\A})^{\epsilon = -1}_{k=4}\big)}$ is spanned by the set of generators $\{U_{_a} , {\tilde U}^{^a}\}$,
which fulfill the following set of non-trivial (anti-)commutation relations \cite{Eghbali3}:
\begin{align}
[U_{_1} , U_{_4}] =& U_{_3},~~~~~~~~~~~~~~~~\{U_{_4} , U_{_4}\} =U_{_2},~~~~~~~~~~~~~~~~[{\tilde U}^{^2} , {\tilde U}^{^3}]= -{\tilde U}^{^4},~~~~~~~~
{\{{\tilde U}^{^3} , {\tilde U}^{^3}\}}=4{\tilde U}^{^1},\nonumber\\
{[U_{_4} , {\tilde U}^{^2}]}=& - U_{_3}-{\tilde U}^{^4},~~~~~\{U_{_4} , {\tilde U}^{^3}\}= - U_{_2}-{\tilde U}^{^1},~~~~~~
{[U_{_1} , {\tilde U}^{^3}]}=-4U_{_3}-{\tilde U}^{^4}. \label{4.37}
\end{align}
From the Manin supertriples in \eqref{4.30}, it follows that $\D=(({\C}^3 + {\A}), {\cal I}_{_{(2|2)}})$ and
${\D}_{_U}={ \big(({\C}^3 + {\A}) ,}$\\${ ({\C}^3 + {\A})^{\epsilon = -1}_{k=4}\big)}$ are isomorphic as Lie superalgebras.
The corresponding isomorphism transformation is given by
\begin{align}
{T_{_1}} =&U_{_1},~~~~~~~~~~~~~~~~~{\tilde T}^{^1} = -U_{_2}+{\tilde U}^{^1},\nonumber\\
{T_{_2}} =&U_{_2},~~~~~~~~~~~~~~~~~{\tilde T}^{^2} =U_{_1}+{\tilde U}^{^2},\nonumber\\
{T_{_3}} =&-U_{_3},~~~~~~~~~~~~~~{\tilde T}^{^3} =-2U_{_4}+ {\tilde U}^{^3},\nonumber\\
{T_{_4}} =&-U_{_4},~~~~~~~~~~~~~~{\tilde T}^{^4} =-2U_{_3} + {\tilde U}^{^4}.\label{4.38}
\end{align}
From the relations \eqref{4.6}-\eqref{4.8}, we obtain the matrices $M_{_{ab}}$, ${N^{^a}}_{_{b}}$ and ${\Pi}^{^{ab}}(g_{_U})$, yielding
\begin{eqnarray}
NM^{^{-1}} = \left( \begin{array}{cccc}
\frac{-1}{1-{\frac{2M}{r}}}  & -1 & 0 & 0\\
1  & \frac{1}{{r^{2} \sin^2 \theta}}  & 0 & 0\\
0 & 0  & \frac{-h(r)}{{F_{0}}^2} & \frac{1}{F_{0}}-2\\
0  & 0  & -\frac{1}{F_{0}}-2& 0\\
\end{array} \right),~~{\Pi}(g_{_U}) = \left( \begin{array}{cccc}
0  & 0  & 0 & 0\\
0  &0  & -\chi & 0\\
0 & \chi  & -4 x_{_{1}} & 0\\
0  & 0  & 0& 0\\
\end{array} \right),~~~\label{4.39}
\end{eqnarray}
then, equation \eqref{4.5} leads to
\begin{eqnarray}
{E}_{_{ab}}(g_{_U}) = \left( \begin{array}{cccc}
\frac{1-{\frac{2M}{r}}}{\Delta} & \frac{\Delta+1}{\Delta}  & 0 & -\frac{F_{_{0}}(\Delta+1)}{(1+2F_{_{0}}) \Delta} \chi \\
-\frac{\Delta+1}{\Delta}  & -\frac{r^{2} \sin^2 \theta}{\Delta}  & 0 & \frac{F_{_{0}} r^{2} \sin^2 \theta}{(1+2F_{_{0}})\Delta} \chi\\
0 & 0  & 0 & \frac{F_{_{0}}}{1+2F_{_{0}}}\\
\frac{F_{_{0}}(\Delta+1)}{(-1+2F_{_{0}})\Delta} \chi  & \frac{F_{_{0}} r^{2} \sin^2 \theta}{(-1+2F_{_{0}})\Delta} \chi  & \frac{F_{_{0}}}{-1+2F_{_{0}}} & -\frac{h(r)+4F_{_{0}}^{2}x_{1}}{4F_{_{0}}^{2}-1}\\
\end{array} \right).\label{4.40}
\end{eqnarray}
Finally, using \eqref{4.4} and \eqref{4.40}, we construct the $\sigma$-model on the superdouble ${ \big(({\C}^3 + {\A}) , ({\C}^3 + {\A})^{\epsilon = -1}_{k=4}\big)}$ whose background is
\begin{align}
ds^2 =& \frac{1}{1-{\frac{2M}{r}}} dr^2 + r^{2} d{\theta}^2 +\frac{1-{\frac{2M}{r}}}{\Delta} d{x_{_{1}}}^2  -\frac{r^{2} \sin^2 (\theta)}{\Delta} d{x_{_{2}}}^2
- \frac{2 F_{_{0}}(\Delta+1)}{(4F_{_{0}}^{2}-1)\Delta} \chi~ d{x_{_{1}}}  d \chi \nonumber \\
& ~~~~~~~~~~~~~~~~~~~~~~~~~~~~~~~~~~~~~~~~~~~~~~~~~~~~~ - \frac{r^{2} \sin^2 \theta}{(4F_{_{0}}^{2}-1)\Delta} \chi~ d{x_{_{2}}}  d \chi + \frac{2F_{_{0}}}{4F_{_{0}}^{2}-1} ~d \psi  d \chi,  \nonumber\\
B=&\frac{\Delta+1}{\Delta}d{x_{_{1}}} \wedge d{x_{_{2}}} + \frac{\Delta+1}{2(4F_{_{0}}^{2}-1)\Delta}\chi ~ d{x_{_{1}}} \wedge  d \chi  \nonumber\\
& ~~~~~ +  \frac{F_{_{0}} r^{2} \sin^2 \theta}{(4F_{_{0}}^{2}-1)\Delta}\chi ~ d{x_{_{2}}} \wedge  d \chi -\frac{2F_{_{0}}^{2}}{4F_{_{0}}^{2}-1} d \psi \wedge  d \chi + \frac{h(r)}{2(4F_{_{0}}^{2}-1)} ~ d \chi \wedge  d \chi.\label{4.41}
\end{align}
To evaluate the total dilaton contribution, we utilize $\phi^{^{(0)}}= {\cal C}_{_{0}}$ and equation \eqref{4.9}, which yields the final result
\begin{eqnarray}
\Phi_{_U} = {\cal C}_{_{0}}+\frac{1}{2}\log\Big|\frac{4F_{_{0}}^{2}-1}{\Delta}\Big|.\label{4.42}
\end{eqnarray}
Consequently, the new background with the aforementioned dilaton satisfies the beta-function equations at the one-loop order.

\section{Conclusion}
\label{Sec.V}

In this paper, we have studied supersymmetric deformations of Schwarzschild spacetime within the super Poisson-Lie T-duality/plurality framework.
The construction was performed on six-dimensional supermanifolds including four bosonic coordinates and two fermionic fields.
By considering the $(2|2)$-dimensional Lie supergroups $(C^{3} + A)$, $C^{3} \otimes A_{1,1}$, and $(2{ A}_{1,1}+2{ A})^0$, a number of the $\sigma$-model backgrounds related to different Lie superalgebraic structures were obtained.
The original backgrounds preserve the Schwarzschild form in their purely bosonic sector,
while the fermionic coordinates lead to non-trivial extension of geometry. The vanishing of the one-loop beta-function equations was used to determine the corresponding dilaton fields and to check the conformal invariance of the obtained models involving the vanishing cosmological constant.

Our initial analysis of super non-Abelian T-duality was performed on semi-Abelian Drinfeld superdoubles.
Though the three Lie supergroups which lie behind them are not isomorphic, a nice structure emerges in the dual models.
Specifically, the three dual metrics corresponding to $(C^{3} + A)$, $C^{3} \otimes A_{1,1}$, and $(2{ A}_{1,1}+2{ A})^0$ have exactly the same form.
Thus, the quantities which depend only on the metric, for example, the scalar curvature or Kretschmann invariant, take identical values for these models.
So the structure of singularities in the dual metrics is common to the three examples.
Nevertheless, the corresponding $B$-fields are different, and the difference between the dual models is preserved both in the fermionic sector and in their coupling to the background fields.
This shows that the information encoded in the underlying Lie superalgebra need not always be visible from the bosonic metric or encoded in curvature invariants.
The superisometry considerations of the original backgrounds give yet another hint about that difference.
In particular, the background corresponding to $(2{ A}_{1,1}+2{ A})^0$ admits five bosonic and two fermionic Killing supervectors.
One of the fermionic generators contains both a fermionic translation and a time-translation component,
which is consistent with the presence of Grassmannian coupling between the time and fermionic coordinates in the metric.
Therefore, the super Killing structure allows to differentiate supergeometries with the same Schwarzschild-like bosonic sector and the same $B$-field.
On the contrary, for three dual models having the same metric, the superisometric properties defined by the metric become the same, while $B$-fields give an additional difference among the backgrounds.

Subsequently, we have applied super Poisson-Lie T-plurality to different Manin supertriple decompositions of the relevant Drinfeld superdoubles.
The super plurality transformations lead to further conformal backgrounds which are not, in general, equivalent
to the backgrounds obtained directly by the super non-Abelian T-duality.
In one of the resulting models, the Schwarzschild bosonic sector is retained and the main modification occurs in the fermionic part and its coupling to the bosonic coordinates.
In another case, the quantity $\Delta$ enters directly into the bosonic metric and the $B$-field, so that the effect of the plurality transformation is no longer restricted to the fermionic sector. The corresponding dilaton fields obtained from the super Poisson-Lie transformation rule ensure that these backgrounds satisfy the one-loop beta-function equations.
For the last plurality background with respect to the $(C^{3} + A)$, the function $\Delta$ is also included in the metric, $B$-field,
and dilaton, providing another example in which different decompositions of a Drinfeld superdouble generate non-trivial deformations of the original Schwarzschild geometry.

The outcomes of this study reveal that the methods of super Poisson-Lie T-duality and T-plurality can be used to systematically construct a class of conformal supergeometries from a Schwarzschild background coupled to fermionic fields.
An important point about these constructions is that algebraically different or differently decomposed superdoubles can lead to backgrounds with identical bosonic sectors, while their fermionic interactions, $B$-fields, dilatons, and superisometry structures retain information about the underlying superalgebra.
In cases where the plurality transformation changes the bosonic sector as well, the resulting backgrounds exhibit a more direct geometric manifestation of the superalgebraic data. A more detailed study of the global properties of these backgrounds, their complete superisometry algebras, and their possible interpretation within generalized supergravity or double field theory would provide natural extensions of the present analysis.

\subsection*{Acknowledgements}

The authors would like to express their gratitude to the handling editor and reviewers for their constructive comments
and helpful suggestions.
\\
\\
{\bf Data Availability Statement.} This article has no associated data or the data will not
be deposited.
\\
\\
{\bf Code Availability Statement.} This article has no associated code or the code will not
be deposited.
\\
\\
{\bf ORCID iDs}
\\
A. Eghbali ~~~~~~~~~~~~~~~~  0000-0001-6076-2179
\\
M. Hosseinpour-Sadid ~ 0009-0000-5418-1896
\\
A. Rezaei-Aghdam ~~~~~ 0000-0003-4754-7911


\end{document}